\documentclass[pra,preprint,showpacs,showkeys,nofootinbib,tightenlines,floatfix,superscriptaddress]{revtex4}

\usepackage{ulem}
\usepackage{graphicx}  %
\usepackage{amsmath}
\usepackage{bm}  %

\newcommand{\bea}{\begin{eqnarray}}
\newcommand{\eea}{\end{eqnarray}}
\newcommand{\beq}{\begin{equation}}
\newcommand{\eeq}{\end{equation}}
\newcommand{\bqa}{\begin{eqnarray}}
\newcommand{\eqa}{\end{eqnarray}}

\def\mqo2{{\!\!\!}}

\begin{document}

\title{Efimov Physics in $\bm{^6}$Li Atoms}

\author{Eric Braaten}
\affiliation{Department of Physics,
         The Ohio State University, Columbus, OH\ 43210, USA}
\affiliation{Helmholtz-Institut f\"ur Strahlen- und Kernphysik
        (Theorie) and Bethe Center for Theoretical Physics,
        Universit\"at Bonn, 53115 Bonn, Germany}

\author{H.-W. Hammer}
\affiliation{Helmholtz-Institut f\"ur Strahlen- und Kernphysik
        (Theorie) and Bethe Center for Theoretical Physics,
        Universit\"at Bonn, 53115 Bonn, Germany}

\author{Daekyoung Kang}
\affiliation{Department of Physics,
         The Ohio State University, Columbus, OH\ 43210, USA}

\author{Lucas Platter}
\affiliation{Department of Physics,
         The Ohio State University, Columbus, OH\ 43210, USA}
\date{August 2009\\[0.5cm]}

\begin{abstract}
  A new narrow 3-atom loss resonance associated with an Efimov trimer crossing
  the 3-atom threshold has recently been discovered in a many-body system of
  ultracold $^6$Li atoms in the three lowest hyperfine spin states at a
  magnetic field near 895~G. O'Hara and coworkers have used measurements of
  the 3-body recombination rate in this region to determine the complex
  3-body parameter associated with Efimov physics.  Using this parameter as
  the input, we calculate the universal predictions for the spectrum of
  Efimov states and for the 3-body recombination rate in the universal region
  above 600~G where all three scattering lengths are large. We predict an
  atom-dimer loss resonance at $672 \pm 2$~G associated with an Efimov
  trimer disappearing through an atom-dimer threshold.  We also predict an
  interference minimum in the 3-body recombination rate at $759 \pm 1$~G
  where the 3-spin mixture may be sufficiently stable to allow experimental
  study of the many-body system.
\end{abstract}

\smallskip
\maketitle

\section{Introduction}
 
Ultracold atomic gases allow the experimental study of many-body physics in
systems in which the fundamental interactions are theoretically understood and
experimentally controllable.  There are many possibilities for the atomic
constituents of the many-body system: they can be identical atoms that are all
in the same spin state or there can be multiple species of constituents that
are either different spin states of the same atom, different isotopes of an
atom, or atoms of different elements.  If the atoms are identical fermions in
the same spin state, the many-body system is essentially an ideal gas.  
If the atoms are fermions with two spin states, there is a well-behaved
zero-range limit in which the interactions are completely determined by the
scattering length $a$.  The ground state of the many-body system 
with equal populations of the two spin states is a superfluid. 
As $a$ is changed from a negative
value to a positive value through $\pm \infty$, the mechanism for
superfluidity changes smoothly from the {\it BCS mechanism} (Cooper pairing of
atoms) to the {\it BEC mechanism} (Bose-Einstein condensation of diatomic
molecules).  In the {\it unitary limit} in which the pair scattering length is
infinitely large, the behavior of the system is constrained by {\it scale
  invariance}: invariance under rescaling space and time by arbitrary positive
numbers $\lambda$ and $\lambda^2$, respectively.  There have been extensive
investigations, both experimental and theoretical, of the universal behavior
of the system in the crossover region \cite{IKS08}.

Fermionic atoms with three spin states open up new possibilities for universal
behavior. In the many-body system, there can be new superfluid phases and new
mechanisms for superfluidity \cite{Bedaque:2006ii,PMT06,HJZ06,Zhai06,MKTP09}. 
The universal behavior can be more complex, because there are three independent
scattering lengths.  However it is also qualitatively different because of the
{\it Efimov effect}.  This remarkable phenomenon was discovered by Vitaly
Efimov in the 3-body problem for identical bosons \cite{Efimov}. 
In the {\it  unitary limit} in which the three pair scattering lengths 
are all infinitely
large, there is an infinite sequence of 3-body bound states called 
{\it Efimov  trimers} with an accumulation point at the 3-atom threshold. 
The ratio of the binding energies of successive Efimov trimers is 
approximately 1/515.  This geometric spectrum reflects 
{\it discrete  scale invariance}: invariance under rescaling space and 
time by the same
integer power of $\lambda_0\approx 22.7$ and $\lambda_0^2\approx 515$, 
respectively. 
Low-energy phenomena associated with discrete scale invariance
are generally referred to as {\it  Efimov physics} 
\cite{Braaten:2004rn,Braaten:2006vd,Platter:2009gz}.  (See
Ref.~\cite{Platter:2009gz} for a summary of recent developments.)  Efimov
physics can also occur in other 3-body systems if at least two of the three
pair scattering lengths are large.  The discrete scaling factor depends on the
mass ratios of the particles and their symmetries.  In the case of fermions
with 3 spin states for which the 3 pair scattering lengths are all large, the
Efimov effect occurs with the same discrete scaling factor $\lambda_0$ 
as for identical bosons \cite{Efimov73}.

The first experimental evidence for Efimov physics in ultracold atoms was
presented by Kraemer et al.\ \cite{Kraemer:2006}. Their atoms were $^{133}$Cs
atoms
in the lowest hyperfine spin state, which are identical bosons.  They observed
resonant enhancement of the loss of atoms from 3-body recombination that can
be attributed to an Efimov trimer crossing the 3-atom threshold. 
The dependence of the three-body recombination rate on the scattering
length was first calculated in Refs.~\cite{NM-99,EGB-99,BBH-00}. In 
particular, Esry, Greene and Burke predicted a resonant enhancement 
of the three-body recombination rate at negative values of the scattering
length~\cite{EGB-99}. The universal
line shape for the resonance as a function of the scattering length was first
derived by Braaten and Hammer \cite{Braaten:2003yc}.  Kraemer et al.\ also
observed a minimum in the 3-body recombination rate that can be interpreted as
an interference effect associated with Efimov physics.  In a subsequent
experiment with a mixture of $^{133}$Cs atoms and dimers, Knoop et al.\
observed a resonant enhancement in the loss of atoms and dimers
\cite{Knoop:2008}.  This loss feature can be explained by an Efimov trimer
crossing the atom-dimer threshold \cite{Helfrich:2009uy}.  The most exciting
recent developments in the Efimov physics of $^{133}$Cs atoms involve
universal tetramer states.  Platter and Hammer discovered that there is a pair
of universal tetramer states associated with every Efimov trimer and they
calculated their binding energies for limited ranges of the scattering length
\cite{Hammer:2006ct}. Von Stecher, D'Incao and Greene mapped out their
spectrum for all scattering lengths and pointed out that resonant enhancement
of 4-body recombination would provide a signature for these tetramers
\cite{Stecher:2008}. Ferlaino et al.\ recently observed 
the loss resonances from both tetramers in an
ultracold gas of $^{133}$Cs atoms \cite{Ferlaino:2009}.

Recent experiments with other bosonic atoms have also provided evidence of
Efimov physics.  Zaccanti et al.\ measured the 3-body recombination rate and
the atom-dimer loss rate in a ultracold gas of $^{39}$K atoms
\cite{Zaccanti:2008}.  They observed two atom-dimer loss resonances and two
minima in the 3-body recombination rate at large positive values of the
scattering length.  The positions of the loss features are consistent with the
universal predictions with discrete scaling factor 22.7.  They also observed
loss features at large negative scattering lengths.  Barontini et al.\
obtained the first evidence of the Efimov effect in a heteronuclear mixture of
$^{41}$K and $^{87}$Rb atoms \cite{Barontini:2009}.  They observed 3-atom loss
resonances at large negative scattering lengths in both the K-Rb-Rb and K-K-Rb
channels, for which the discrete scaling factors are 131 and $3.51 \times
10^5$, respectively.  Gross et al.\ measured the 3-body recombination
rate in an ultracold system of $^7$Li atoms \cite{Gross:2009}.  They observed
a 3-atom loss resonance at a large negative scattering length and a 3-body
recombination minimum at a large positive scattering length.  The positions of
the loss features, which are in the same universal region on different sides
of a Feshbach resonance, are consistent with the universal predictions with
discrete scaling factor 22.7.

A promising fermionic atom in which to observe Efimov physics is $^6$Li.  For
the three lowest hyperfine states of $^6$Li atoms, the three pair scattering
lengths approach a common large negative value at large magnetic fields and
all three have nearby Feshbach resonances at lower fields that can be used to
vary the scattering lengths \cite{Bartenstein:2005}.  The first experimental
studies of many-body systems of $^6$Li atoms in the three lowest hyperfine
states have recently been carried out by Ottenstein et al.\
\cite{Ottenstein:2008} and by Huckans et al.\ \cite{Huckans:2008fq}.  Their
measurements of the 3-body recombination rate revealed a narrow loss feature
and a broad loss feature in a region of low magnetic field.  Theoretical
calculations of the 3-body recombination rate supported the interpretation
that the narrow loss feature arises from an Efimov trimer crossing the 3-atom
threshold \cite{Braaten:2008wd,NU:2009,Schmidt:2008fz}.  Very recently,
another narrow loss feature was discovered in a much higher region of the
magnetic field by Williams et al.\ \cite{Williams:2009} and by Jochim and
coworkers \cite{Jochim}. Williams et al.\ used measurements of the 3-body
recombination rate in this region to determine the complex 3-body parameter
that governs Efimov physics in this system.  This parameter, together with
the three scattering lengths as functions of the magnetic field, determine the
universal predictions for $^6$Li atoms in this region of the magnetic field.

In this paper, we calculate universal predictions for various aspects of
Efimov physics for the three lowest hyperfine spin states of $^6$Li atoms.  In
Section~\ref{sec:theory}, we explain how universal predictions for 3-body
observables can be calculated efficiently by solving coupled sets of integral
equations. We apply these methods specifically to the 3-body recombination
rate and to the binding energies and widths of Efimov trimers. In
Section~\ref{sec:lofield}, we summarize previous experimental and theoretical
work on $^6$Li atoms in the universal region at low magnetic fields where
all 3 scattering lengths are negative and relatively large. In
Section~\ref{sec:hifield}, we use the complex 3-body parameter determined by
Williams et al.\ to calculate universal predictions for the binding energies
and widths of Efimov trimers and for the three-body recombination rate in
regions where one or more of the scattering lengths are large and positive.
We predict an atom-dimer resonance at $672 \pm 2$~G where an Efimov trimer
disappears through an atom-dimer threshold.  We predict an interference
minimum in the 3-body recombination rate at $759 \pm 1$~G where the 3-spin
mixture may be sufficiently stable to allow experimental study of the
many-body system. We also discuss the implications of our predictions for the
many-body physics of $^6$Li atoms. We summarize our results in
Section~\ref{sec:discussion}.

\section{Theoretical Formalism}
\label{sec:theory}

\subsection{Fermions with Three Spin States}
\label{sec:3fermions}

We consider a fermionic atom of mass $m$ with three spin states that we label
as types 1, 2, and 3.  We denote the scattering length of the pair $ij$ by
either $a_{ij} = a_{ji}$ or $a_k$, where $(ijk)$ is a permutation of $(123)$.
We assume that the three pair scattering lengths $a_{12}= a_3$, $a_{23} =
a_1$, and $a_{13}= a_2$ are all much larger than the range of interactions.
The two-body physics for fermions with two distinct spin states that have a
large pair scattering length $a_{ij}$ is very simple in the zero-range limit.
The scattering amplitude for the pair $ij$ with relative wavenumber $q$ is
$(-1/a_{ij} - iq)^{-1}$.  If $a_{ij}$ is positive, there is a weakly-bound
diatomic molecule with constituents $i$ and $j$ and with binding energy
$\hbar^2/(m a_{ij}^2)$.  We will refer to it as the {\it $(ij)$ dimer}.  We
refer to the (12), (23), and (13) dimers collectively as {\it shallow dimers}.
For realistic interactions with finite range $R$, there can also be 
deeply-bound
diatomic molecules whose binding energies are comparable to or larger than the
energy scale $\hbar^2/(m R^2)$ set by the range.  We refer to them as 
{\it  deep dimers}.  Their binding energies are insensitive to changes in the
large scattering lengths.

For the 3-atom problem, we take the zero of energy to be the scattering
threshold for the three atoms. If $a_{jk}$ is positive, the scattering
threshold for an atom of type $i$ and a $(jk)$ dimer is 
$-\hbar^2/(m a_{jk}^2)$.
We will refer to this scattering threshold as the 
{\it $i+(jk)$ atom-dimer threshold}.
We will refer to triatomic molecules as {\it trimers}.
A trimer whose constituents are atoms of types 1, 2, and 3
must have energy below the 3-atom threshold and below the 
$i+(jk)$ atom-dimer threshold if $a_{jk} > 0$.

The three-body physics for fermions with three distinct spin states is more
complicated than for two distinct spin states, because now there are three
scattering lengths $a_{12}$, $a_{23}$, and $a_{13}$.  Depending on the signs
of these scattering lengths, there can be 0, 1, 2, or 3 types of shallow
dimers.  The three-body physics is also much more intricate due to the Efimov
effect. The unitary limit in which all three scattering
lengths are infinitely large is characterized by discrete scale invariance
with discrete scaling factor $\lambda_0 = e^{\pi/s_0} \approx 22.7$, 
where  $s_0 \approx 1.00624$ is a transcendental number.  This requires a
3-body parameter that provides another length scale in addition to the three
scattering lengths. If there were no such parameter, the system would have
continuous scale invariance in the unitary limit.

One simple choice for the 3-body parameter is a wavenumber $\kappa_*$ defined
by the spectrum of 3-body bound states ({\it Efimov trimers}) in the unitary
limit in which all three scattering lengths are infinite \cite{Braaten:2004rn}.
Their binding energies relative to the 3-atom threshold are
\begin{equation}                               
E_T^{(n)} = \lambda_0^{2(n_*-n)} \frac{\hbar^2 \kappa^2_*}{m}
 ~~~~~~~~~~ (a_{12} = a_{23} = a_{31} = \pm \infty)\,,
\label{Efimov}
\end{equation}
where $\kappa_*$ is the binding wavenumber of the trimer labeled $n_*$
and $n$ is an integer.  There are infinitely many of these Efimov trimers
with an accumulation point at the threshold for three atoms of types 1,
2, and 3. The advantage of using the Efimov
parameter $\kappa_*$ as the 3-body parameter is that physical observables
must be log-periodic functions of $\kappa_*$.
In the zero-range limit, the Efimov spectrum also extends infinitely deep
to $-\infty$.
For finite range $R$, the deepest Efimov trimer may have a binding
energy $E_T$ as large as $\hbar^2/(m R^2)$.
However, universal predictions will be accurate only
if its binding energy is significantly smaller,
because range corrections are expected
to be of order $R (mE_T/\hbar^2)^{1/2}$.

In the unitary limit, the Efimov states are sharp states with the spectrum in
Eq.~(\ref{Efimov}) only if there are no deep dimers in any of the three 
2-body channels.  If there are deep dimers, their inclusive effects can
be taken into account by analytically continuing the Efimov parameter
$\kappa_*$ to a complex value that is conveniently expressed in the form
$\kappa_* \exp (i \eta_* / s_0)$, where $\kappa_*$ and $\eta_*$ are positive
real parameters \cite{Braaten:2003yc}.  Making the substitution $\kappa_* \to
\kappa_* \exp (i \eta_* / s_0)$ on the right side of Eq.~(\ref{Efimov}), we
find that the Efimov trimers acquire nonzero decay widths $\Gamma_T^{(n)}$.
The binding energies and widths are
\begin{eqnarray}                               
E_T^{(n)} &=& 
 \lambda_0^{2(n_*-n)} 
\frac{\hbar^2 \kappa_*^2 \cos(2\eta_*/s_0)}{m} 
 ~~~~~~~~~~ (a_{12} = a_{23} = a_{31} = \pm \infty)\,,
\\
\Gamma_T^{(n)} &=& 
\lambda_0^{2(n_*-n)} 
\frac{2 \hbar^2 \kappa_*^2 \sin(2\eta_*/s_0)}{m} \,.
\label{Efimov-eta}
\end{eqnarray}

\subsection{Three-body Recombination}
\label{sec:3bodyrecomb}

Three-body recombination is a three-atom collision in which two of the atoms
form a dimer. In the case of three fermions in the same spin state, 3-body
recombination is strongly suppressed at low temperature, because each pair of
atoms has only P-wave interactions. In the case of two fermions of type $i$
and a third atom of a different type $j$, two of the pairs have S-wave
interactions with scattering length $a_{ij}$. The rate for 3-body
recombination in the zero-range limit still decreases to 0 as the energy $E$
of the atoms approaches the threshold, decreasing like $E$ if $a_{ij} > 0$
\cite{Petrov03} and like $E^3$ if $a_{ij} < 0$ \cite{DIncao05}. In the case
of three distinct spin states, all three pairs of atoms can have S-wave
interactions.  There is no threshold suppression 
of three-body recombination if at least
two of the three scattering lengths are large.  If $a_{ij}$ is large and
positive, one of the recombination channels is into the $(ij)$ dimer and a
recoiling atom with complimentary spin $k$.  If there are deep dimers in any
of the three 2-body channels, they provide additional recombination channels.
If all three scattering lengths are negative, the only recombination channels
are into deep dimers.

The rate equations for the number densities $n_i$ of atoms in the three spin
states are
\begin{equation}
\frac{d\ }{dt} n_i = - K_3 n_1 n_2 n_3.
\label{dndt}
\end{equation}
The event rate constant $K_3$ can be separated into the inclusive rate
constant $K_3^{\rm deep}$ for recombination into deep dimers and the exclusive
rate constants $K_3^{(ij)}$ for recombination into each of the three possible
shallow dimers:
\begin{eqnarray}
K_3 &=& 
K_3^{\rm deep} + K_3^{\rm shallow} ,
\label{K3-total}
\\
K_3^{\rm shallow} &=& 
K_3^{(12)} + K_3^{(23)} + K_3^{(13)} .
\label{K3-shallow}
\end{eqnarray}
The term $K_3^{(ij)}$ is nonzero only if $a_{ij} > 0$.

In the low-temperature limit, the rate constant $K_3$ and the exclusive rate
constants $K_3^{(ij)}$ can be expressed in terms of T-matrix elements for
processes in which the initial state consists of three atoms in the spin
states 1, 2, and 3 with momentum $\bm{0}$.  By the optical theorem, $K_3$ is
twice the imaginary part of the forward T-matrix element for 3-atom elastic
scattering in the limit where the momenta of the atoms all go to 0:
\begin{equation}
K_3 = 
2 \, {\rm Im} \, {\cal T}(\bm{0},\bm{0},\bm{0};\bm{0},\bm{0},\bm{0}) .
\label{K3tot-T}
\end{equation}
The T-matrix element is singular as all the momenta go to zero, but its
imaginary part is not.  If $a_{ij} > 0$, the rate constant $K_3^{(ij)}$ for
recombination into the $(ij)$ dimer is the square of the T-matrix element for
three atoms with momentum $\bm{0}$ to scatter into the dimer and a recoiling
atom multiplied by the atom-dimer phase space:
\begin{equation}
K_3^{(ij)} = 
\frac{4 m}{3 \sqrt{3} \pi \hbar a_k} 
\left| {\cal T}_k(\bm{0},\bm{0},\bm{0};\bm{p},-\bm{p}) \right|^2
\bigg|_{|\bm{p}| = 2\hbar/(\sqrt{3} a_k)} .
\label{K3ij-T}
\end{equation}
The dimer and the recoiling atom with complementary spin $k$ both have
momentum $2\hbar/(\sqrt{3} a_k)$. For convenience, we will switch to
wavenumber variables in the remainder of the paper.

The T-matrix elements in Eqs.~(\ref{K3tot-T}) and (\ref{K3ij-T}) can be
expressed in terms of amplitudes ${\cal A}_{ij}(p,q;E)$ for the transition
from an atom of type $i$ and a complimentary diatom pair
into an atom of type $j$ and a complimentary diatom pair,
with the two diatom pairs being the first to interact and the last 
to interact, respectively.
The projection onto S-waves reduces the amplitude to a function of three
variables: the relative wavenumber $p$ of the incoming atom and diatom, the
relative wavenumber $q$ of the outgoing atom and diatom, and the total energy
$E$ of either the incoming atom and diatom or the outgoing atom and diatom.
The rate constant $K_3$ in Eq.~(\ref{K3tot-T}) can be expressed as
\begin{equation}                               
K_3  = \frac{32 \pi^2\hbar}{m} \sum_{i,j} a_i a_j {\rm Im} A_{ij}(0,0;0),
\label{K3total-A}
\end{equation}
where the sums are over
$i,j = 1,2,3$.  The exclusive rate constant in Eq.~(\ref{K3ij-T}) for 3-body
recombination into the $(ij)$ dimer can be expressed as
\begin{equation}
K_3^{(ij)} =
\frac{512 \pi^2 \hbar}{3 \sqrt{3} m a_k^2} 
\left| \sum_l a_l {\cal A}_{lk}(0,2/(\sqrt{3} a_k);0) \right|^2,
\label{K3ij-A}
\end{equation}
where $k$ is the complimentary spin to $ij$ and the sum is over 
$l = 1,2,3$.

\subsection{STM Equations}
\label{sec:STMeq}

The 9 amplitudes ${\cal A}_{ij}(p,q;E)$ satisfy coupled integral equations in
the variable $q$ that are generalizations of the Skorniakov--Ter-Martirosian
(STM) equation \cite{STM57}. To determine the rate constants for 3-body
recombination in Eqs.~(\ref{K3total-A}) and (\ref{K3ij-A}), it is sufficient
to set the relative wavenumber 
in the initial state to 0 and the total energy to
0. The 9 coupled STM equations for ${\cal A}_{ij} (0,p;0)$ are 
\cite{Braaten:2008wd}
\begin{eqnarray}
{\cal A}_{ij}(0,p;0) &=& \frac{1 - \delta_{ij}}{p^2}
+ \frac{2}{\pi} \sum_k (1 - \delta_{kj}) 
\int_0^\Lambda \! dq \, 
Q(p,q;0) D_k(3q^2/4) {\cal A}_{ik}(0,q;0) ,
\label{STMeq}
\end{eqnarray}
where 
\begin{eqnarray}
Q(p,q;E) &=& 
\frac{q}{2p} \log \frac{p^2 + pq + q^2-mE/\hbar^2}{p^2 - pq + q^2-mE/\hbar^2} ,
\\
D_k(p^2) &=& \left[ -1/a_k + \sqrt{p^2 - i \epsilon} \right]^{-1},
\label{Dk}
\end{eqnarray}
and $\Lambda$ is an ultraviolet cutoff that must be large compared
to $p$, $1/|a_1|$, $1/|a_2|$, and $1/|a_3|$.  Since the T-matrix elements in
Eqs.~(\ref{K3total-A}) and (\ref{K3ij-A}) involve only the three linear
combinations $\sum_i a_i {\cal A}_{ij}(0,p;0)$, the 9 coupled STM equations
can be reduced to 3 coupled integral equations for these 3 linear
combinations. If $\Lambda$ is sufficiently large, the solutions to the
integral equations in Eqs.~(\ref{STMeq}) depend only log-periodically on
$\Lambda$ with a discrete scaling factor $\lambda_0 \approx 22.7$. The
dependence on the arbitrary cutoff $\Lambda$ can be eliminated in favor of a
physical 3-body parameter, such as the Efimov parameter $\kappa_*$ defined by
Eq.~(\ref{Efimov}). If we restrict $\Lambda$ to a range that corresponds to a
multiplicative factor of 22.7, then $\Lambda$ differs from $\kappa_*$ only by
a multiplicative numerical constant. Thus, we can also simply use $\Lambda$ as
the 3-body parameter \cite{Hammer:2000nf}.

If $\Lambda$ is real valued, the STM equations describe atoms that have no
deep dimers. The rate $K_3^{\rm deep}$ for 3-body recombination into deep
dimers, which can be obtained by combining Eqs.~(\ref{K3-total}),
(\ref{K3-shallow}), (\ref{K3total-A}), and (\ref{K3ij-A}), must therefore be
zero. For atoms that have deep dimers, the effects of the deep dimers can be
described indirectly by using a complex-valued 3-body parameter.  If we use
the ultraviolet cutoff $\Lambda$ as the 3-body parameter, the inclusive
effects of deep dimers can be taken into account by analytically continuing
the upper endpoint of the integral $\Lambda$ in Eqs.~(\ref{STMeq}) to a
complex value $\Lambda \exp (i \eta_* / s_0)$.  The path of integration in the
variable $q$ can be taken to run along the real axis from 0 to $\Lambda$ and
then along the arc from $\Lambda$ to $\Lambda \exp (i \eta_* / s_0)$.  This
path can be deformed to run along the straight line from 0 to $\Lambda \exp (i
\eta_* / s_0)$ provided we add explicitly the contributions from any poles
that are crossed as the contour is deformed.  These poles arise from the
diatom propagator $D_k(p^2)$, which in the case $a_k > 0$ has a pole
associated with the shallow dimer at $p = 1/a_k$.  For example, if $a_k > 0$,
the integral in Eq.~(\ref{STMeq}) with a complex ultraviolet cutoff $\Lambda
e^{i \eta_*/s_0}$ can be written
\begin{eqnarray}
&& \left( \int_0^{\Lambda} \! dq 
         + \int_\Lambda^{\Lambda e^{i \eta_*/s_0}} \!\!\!\! dq \right)
Q(p,q;0) D_k(3q^2/4) {\cal A}_{ik}(0,q;0) 
\nonumber 
\\
&& = Q(p,q_k;0) \frac{4 \pi i}{\sqrt{3}} {\cal A}_{ik}(0,q_k;0) 
+ \int_0^{\Lambda e^{i \eta_*/s_0}} \!\!\!\!\! dq \, 
Q(p,q;0) D_k(3q^2/4) {\cal A}_{ik}(0,q;0) ,
\label{intpole}
\end{eqnarray}
where $q_k = 2/(\sqrt{3} a_k)$. In the integral on the right side,
the integration contour runs along the straight line path from 0 to 
$\Lambda e^{i \eta_*/s_0}$.  With the ultraviolet cutoff replaced by 
$\Lambda \exp (i \eta_* / s_0)$, 
the rate $K_3^{\rm deep}$ for 3-body recombination into deep
dimers is nonzero.

The rate constant $K_3$ in Eq.~(\ref{K3total-A}) requires the extrapolation of
the solutions ${\cal A}_{ij}(0,p;0)$ to the STM equations in
Eqs.~(\ref{STMeq}) to $p=0$. These solutions are singular as $p \to 0$. The
singular terms, which are proportional to $1/p^2$, $1/p$, and $\ln p$, appear
only in ${\rm Re} \, {\cal A}_{ij}(0,p;0)$ for real $p$ and can be derived by
iterating the integral equations \cite{Braaten:2001ay}. Since ${\rm Im} \,
{\cal A}_{ij}(0,p;0)$ must be extrapolated to $p=0$, it is useful to transform
Eqs.~(\ref{STMeq}) into coupled STM equations for amplitudes 
$\bar {\cal  A}_{ij}(0,p;0)$ obtained by subtracting the singular terms from 
${\cal  A}_{ij}(0,p;0)$:
\begin{eqnarray}
\bar {\cal A}_{ij}(0,p;0) = {\cal A}_{ij}(0,p;0)
&-&
\frac{1 - \delta_{ij}}{p^2}
+ \frac{\pi}{3 p} \sum_{n } a_{n} (1-\delta_{in})(1-\delta_{nj})
\nonumber
\\
&-&
\log \frac {p}{\Lambda}
\left\{ \frac{\sqrt{3}}{\pi}\sum_{n}  a_n^2(1-\delta_{in})(1-\delta_{nj})
\right.
\nonumber\\
&&
\left.
-\frac{2}{3}\sum_{m,n} a_m a_n(1-\delta_{im})(1-\delta_{mn})(1-\delta_{nj})
\right\}.
\end{eqnarray}

\subsection{Three Equal Scattering Lengths}
\label{sec:3equala}

We can obtain analytic results for the 3-body recombination rate in
the case of three equal scattering lengths: $a_{12} = a_{23}
= a_{13} = a$.  In this case, the 3-body recombination rates in
Eqs.~(\ref{K3total-A}) and (\ref{K3ij-A}) depend only on the combination
$\sum_i {\cal A}_{ij}(0,p;0)$. One can show that the solutions to the STM
equation for $\sum_i {\cal A}_{ij}(p,q;E)$ are the same for $j = 1,2,3$ and
are equal to the corresponding amplitude ${\cal A}(p,q;E)$ for identical
bosons with scattering length $a$:
\begin{equation}                               
\sum_i {\cal A}_{ij}(p,q;E) = {\cal A}(p,q;E),
~~~~~j=1,2,3.
\end{equation}
The STM equation analogous to Eq.~(\ref{STMeq}) for identical bosons is%
\footnote{The amplitude ${\cal A}(p,q;E)$ differs from the 
amplitude ${\cal A}_S(p,q;E)$ in Ref.~\cite{Braaten:2004rn}
by a multiplicative factor of $a/(8 \pi)$.}

\begin{eqnarray}
{\cal A}(0,p;0) &=& \frac{2}{p^2}
+ \frac{4}{\pi} \int_0^\Lambda \! dq \, 
Q(p,q;0) D(3q^2/4) {\cal A}(0,q;0).
\label{STMeq-boson}
\end{eqnarray}
If the low-temperature limit is taken with the number density for the
identical bosons much smaller than the critical density for Bose-Einstein
condensation, the rate constant for 3-body recombination of identical bosons
into the shallow dimer is \cite{Braaten:2004rn}
\begin{equation}
K_3^{\rm shallow} =
\frac{512 \pi^2 \hbar}{\sqrt{3} m} 
\left| {\cal A}(0,2/(\sqrt{3} a);0) \right|^2.
\label{K3-A:boson}
\end{equation}
Upon making the substitution $\sum_l a_l {\cal A}_{lk} \to a {\cal A}$ in
Eq.~(\ref{K3ij-A}), we see that the recombination rate $K_3^{(ij)}$ for the
three fermions into the $(ij)$ dimer is exactly 1/3 of the recombination rate
for identical bosons in Eq.~(\ref{K3-A:boson}). 
Summing over the three shallow dimers, we find that
the expression for the recombination rate 
of the three fermions into shallow
dimers is identical to the expression for the 
recombination rate of identical bosons
into the single shallow dimer.  Similarly, we find that the expression in
Eq.~(\ref{K3total-A}) for the total recombination rate of the three fermions
is identical to that for the total recombination rate of identical bosons.

Braaten and Hammer deduced a semi-analytic expression for the rate constant
for 3-body recombination of identical bosons into deep dimers with a large
negative scattering length $a$~\cite{Braaten:2003yc}:
\beq
K_3^{\rm deep} =
\frac{16 \pi^2 C \sinh(2 \eta_*)}
    {\sin^2[s_0 \ln(a/a'_*)] + \sinh^2 \eta_*}
\, \frac{\hbar a^4}{m} ~~~~~~ (a<0),
\label{K3-equal}
\eeq
where $s_0 = 1.00624$, $a'_* = -1/(D \kappa_*)$, and $C$ and $D$ are
numerical constants. The most accurate values for the numerical constants are
$C = 29.62(1)$ and $D = 0.6642(2)$ \cite{Braaten:2008wd,Gogolin:2008}.  
The relation
between the Efimov parameter and the ultraviolet cutoff is $\kappa_* =
0.17609(5) \Lambda$, modulo multiplication by an integer power of $\lambda_0
\approx 22.7$.  The expression for the recombination rate constant in
Eq.~(\ref{K3-equal}) exhibits resonant enhancement for $a$ near the values
$\lambda_0^n a_*´$ for which there is an Efimov trimer at the 3-body
threshold.  The line shape in Eq.~(\ref{K3-equal}) 
for the 3-atom loss resonance as a function of the scattering length 
played a key role in the discovery of an Efimov state for
$^{133}$Cs atoms in the lowest hyperfine state \cite{Kraemer:2006}.  It
applies equally well to three fermions with equal negative pair scattering
lengths.

Macek, Ovchinnikov, and Gasaneo \cite{Macek:2005} and Petrov \cite{Petrov-octs}
have deduced a completely analytic expression for the 3-body recombination
rate constant for identical bosons with a large positive scattering length $a$
in the case where there are no deep dimers.  Braaten and Hammer generalized
their result to the case where there are deep dimers by making the analytic
continuation $\kappa_* \to \kappa_* \exp (i \eta_* / s_0)$ in the amplitude
for this process \cite{Braaten:2006vd}.  The resulting analytic expression for
the recombination rate is \cite{Braaten:2006vd}
\begin{eqnarray}
K_3^{\rm shallow} = 
\frac{128 \pi^2 (4 \pi - 3 \sqrt{3})
     (\sin^2 [s_0 \ln (a/a_{*0})] + \sinh^2\eta_*)}
    {\sinh^2(\pi s_0 + \eta_*) + \cos^2 [s_0 \ln (a/a_{*0})] }\,
\frac{\hbar a^4}{m}  ~~~~~~ (a>0)\,,
\label{alpha-analytic:eta}
\end{eqnarray}
%
where $a_{*0} \approx 0.32 \, \kappa_*^{-1}$. This expression exhibits minima for
$a$ near the values $\lambda_0^n a_{*0}$ arising from destructive
interference between two pathways for recombination.  This formula applies
equally well to three fermions with equal positive pair scattering lengths.
It gives the recombination rate $K_3^{\rm shallow}$ in
Eq.~(\ref{K3-shallow}), which is summed over the 3 shallow dimers.

\subsection{Efimov Trimers}
\label{sec:EfTri}

The transition amplitudes ${\cal A}_{ij}(p,q;E)$ have poles 
in the total energy $E$
at the energies $E^{(n)}$ of the Efimov trimers.
Near the pole, the amplitudes factor:
\begin{equation}                               
{\cal A}_{ij}(p,q;E) \longrightarrow 
\frac{{\cal B}_i(p)^* {\cal B}_j(q)}{E - E^{(n)}}.
\label{A-B}
\end{equation}
The spectrum of Efimov trimers can be obtained by solving the
three coupled homogeneous integral equations for ${\cal B}_j(q)$:
\begin{eqnarray}
{\cal B}_j(p) = \frac{2}{\pi} \sum_k (1 - \delta_{kj}) 
\int_0^\Lambda \! dq \, 
Q(p,q;E)  D_k(3q^2/4 - m E/\hbar^2) {\cal B}_{k}(q) ,
\label{STMeqB}
\end{eqnarray}
where the ultraviolet cutoff $\Lambda$ must be much larger than $p$,
$|mE/\hbar^2|^{1/2}$, 
and all three inverse scattering lengths $1/a_i$.  The set of
homogeneous STM equations in Eq.~(\ref{STMeqB}) are nonlinear eigenvalue
equations for $E$.  If $\Lambda$ is real valued, the eigenvalues are real
valued if the energy is below all the scattering thresholds.  The Efimov
trimers are therefore sharp states with 0 widths.  There is always a 3-atom
scattering threshold at $E=0$.  If $a_{ij} > 0$, the atom-dimer scattering
threshold at $E = -\hbar^2/(m a_{ij}^2)$ has lower energy.

If there are deep dimers, the Efimov trimers can decay into a deep dimer and a
recoiling atom.  Their widths can be calculated by analytically continuing the
upper endpoint of the integral $\Lambda$ in Eqs.~(\ref{STMeqB}) to a complex
value $\Lambda \exp (i \eta_* / s_0)$. The complex energy eigenvalue for the
trimer can be expressed as
\begin{equation}                               
E^{(n)} = - E_T^{(n)} - i \Gamma_T^{(n)}/2 .
\label{EGamma}
\end{equation}
If $\Gamma_T^{(n)}$ is small compared to the difference between $E_T^{(n)}$
and the nearest scattering threshold, then
$E_T^{(n)}$ and $\Gamma_T^{(n)}$ can be interpreted as the binding energy 
and the width of the trimer, respectively.  If $\Gamma_T^{(n)}$ is not 
small, they do not have such precise interpretations.

If $\eta_* \ll 1$, the binding energies and widths of the Efimov trimers can be 
calculated approximately by solving the STM equation with a real-valued cutoff 
$\Lambda$.  The complex energy $E^{(n)}$ in Eq.~(\ref{EGamma})
is a function of the complex parameter $\Lambda  \exp(i \eta_*/s_0)$
that must be real valued in the limit $\eta_* \to 0$.  
Expanding that function in powers of $\eta_*$, 
we find that the leading approximations to the binding energy and 
the width are
\begin{eqnarray}                               
E_T^{(n)} &\approx& 
- E^{(n)}\big|_{\eta_* = 0},
\label{E-approx}
\\
\Gamma_T^{(n)} &\approx& 
- \frac{2 \eta_*}{s_0} \Lambda \frac{\partial \ }{\partial \Lambda} 
E^{(n)}\bigg|_{\eta_* = 0}.
\label{Gamma-approx}
\end{eqnarray}
The derivative with respect to $\Lambda$ in Eq.~(\ref{Gamma-approx})
is calculated with the
scattering lengths $a_{ij}$ held fixed.  The leading corrections to
Eqs.~(\ref{E-approx}) and (\ref{Gamma-approx}) are suppressed 
by a factor of $\eta_*^2$.

\subsection{Dimer Relaxation}
\label{sec:ADRelax}

If our system of fermionic atoms is a mixture of $(jk)$ dimers and atoms 
of type $i$, another 3-atom loss process is dimer relaxation:
the inelastic scattering of the atom and the $(jk)$ dimer
into an atom and a dimer with a larger binding energy.
If $i$ coincides with $j$ so there are only two distinct spin states,
the dimer relaxation rate decreases as $a_{ik}^{-3.33}$
as the scattering length increases \cite{Petrov04}.
The atom-dimer mixture is therefore remarkably stable when the scattering 
length is very large.
This stability was verified in the recent experiments with $^6$Li atoms
\cite{Ottenstein:2008,Huckans:2008fq}.
If $i$ is distinct from both $j$ and $k$ so there are three distinct spin 
states, there is no Pauli suppression of the atom-dimer relaxation rate.
The rate increases as the scattering lengths are increased.
On top of that, there can also be resonant enhancement associated with 
Efimov physics.

To be definite, we consider a mixture of (23) dimers and atoms 
of type 1. We denote the number densities of atoms and (23) dimers
by $n_1$ and $n_{(23)}$, respectively.
The loss rate from atom-dimer relaxation can be expressed in terms 
of a rate constant $\beta_{1(23)}$:
\begin{equation}
\frac{d \ }{dt} n_1 = 
\frac{d \ }{dt} n_{(23)} = 
- \beta_{1(23)} n_1  n_{(23)} .
\label{beta-def}
\end{equation}
Atom-dimer relaxation channels are also inelastic atom-dimer scattering 
channels.  So if $\beta_{1(23)} >0$, the atom-dimer scattering length 
$a_{1(23)}$ must have a negative imaginary part.
These two quantities are related by the optical theorem:
\begin{equation}
\beta_{1(23)} = - \frac{6 \pi \hbar}{m} \, {\rm Im} (a_{1(23)}) .
\label{beta-opt}
\end{equation}
If the scattering lengths are all large, then by dimensional 
analysis, $\beta_{1(23)}$ must be $\hbar a_{23}/m$ multiplied by a 
dimensionless coefficient that depends on the ratios of scattering lengths 
$a_{12}/a_{23}$ and $a_{13}/a_{23}$ and also on $a_{23} \kappa_*$, 
where $\kappa_*$ is the Efimov parameter.  The dependence on 
$a_{23} \kappa_*$ is required to be
log-periodic with discrete scaling factor $\lambda_0 \approx 22.7$.
The universal predictions for dimer relaxation rates can be calculated 
by solving appropriate sets of coupled STM equations.
The dimensionless coefficient of $\hbar a_{23}/m$ in 
the relaxation rate constant $\beta_{1(23)}$
can be especially large if there is an Efimov trimer 
close to the 1+(23) atom-dimer threshold. In this case,
there is resonant enhancement of the dimer relaxation rate.
The resulting loss feature is called an {\it atom-dimer loss resonance}.

\section{$\bm{^6}$Li Atoms: Low-field Universal Region}
\label{sec:lofield}

In this section, we apply our formalism to 
the lowest three hyperfine spin states of $^6$Li atoms in
the region of low magnetic field from 0 to 600~G. 

\subsection{Lowest Hyperfine Spin States of $\bm{^6}$Li Atoms}

An example of a fermion with three spin states is $^6$Li atoms 
in the three lowest hyperfine spin states.  They can be labelled 
by their hyperfine quantum numbers $|f, m_f \rangle$
or by integers:
$| 1 \rangle = |\frac12, +\frac12 \rangle$,
$| 2 \rangle = |\frac12, -\frac12 \rangle$, and
$| 3 \rangle = |\frac32, -\frac32 \rangle$.
The pair scattering lengths $a_{12}$, $a_{23}$, and $a_{13}$
have Feshbach resonances near 834~G, 811~G, and 690~G, 
respectively \cite{Bartenstein:2005}.  
Beyond these Feshbach resonances, all three scattering lengths 
approach the triplet scattering length $-2140~a_0$,
which is large and negative.
A convenient conversion constant for $^6$Li atoms is
$\hbar/m = 1.0558 \times 10^{-4}$~cm$^2$/s.

The relevant range for ultracold atoms 
is the van der Waals length $\ell_{\rm vdW} = (m C_6/\hbar^2)^{1/4}$, 
which is approximately $62.5~a_0$ for $^6$Li.
The corresponding energy scale is the 
van der Waals energy $E_{\textrm{vdW}} = \hbar^2/(m \ell_{\textrm{vdW}}^2)$.
The associated frequency is
$\nu_{\textrm{vdW}} = E_{\textrm{vdW}}/(2 \pi \hbar) 
                    \approx 154~{\textrm{MHz}}$.
The zero-range approximation 
should be accurate if $|a_{12}|$, $|a_{23}|$, and $|a_{13}|$
are all much larger than $\ell_{\rm vdW}$ 
and it should be at least qualitatively useful
if $| a_{ij}| > 2 \ell_{\rm vdW}$.
There are two regions of the magnetic field in which all 
three scattering lengths are larger than
$2 \ell_{\rm vdW} \approx 125~a_0$:
a low-field region $122~{\rm G} < B < 485~{\rm G}$
and a high-field region $B > 608~{\rm G}$.
These two universal regions are separated by a nonuniversal 
region in which all three scattering lengths go through zeros.
Efimov physics in the universal regions will be characterized 
by values of $\kappa_*$ and $\eta_*$ that may not be the same 
in the two regions. In general, these parameters may be
expected to vary slowly with the magnetic field, 
just like the scattering length away from a Feshbach resonance.
In a sufficiently narrow region of magnetic field, 
they can be treated as constants.  While their values could 
in principle be calculated from microscopic atomic physics,
in practice they have to be determined by measurements of 
3-body observables.

\begin{figure}[t]
\centerline{\includegraphics*[scale=0.7,angle=0,clip=true]{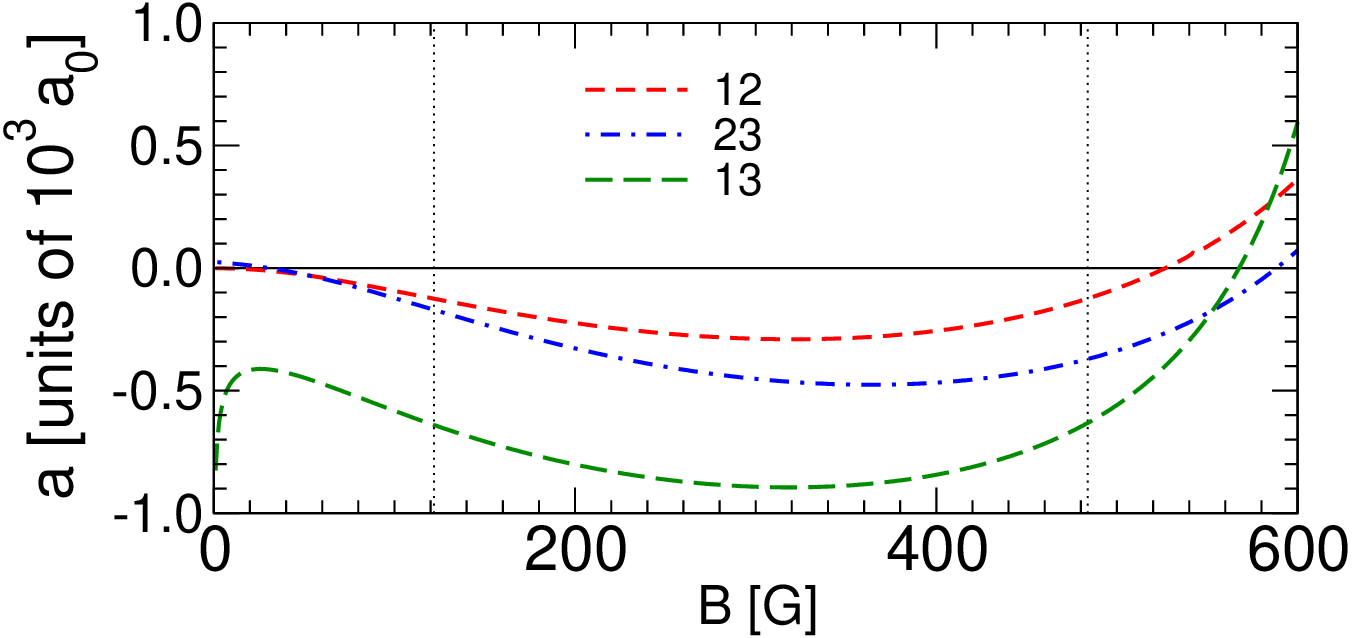}}
\vspace*{0.0cm}
\caption{(Color online) 
The scattering lengths in units of $10^3 a_0$
for the three lowest hyperfine states 
of $^6$Li as functions of the magnetic field $B$
from 0 to 600~G \cite{Julienne}.
The two vertical dotted lines mark the boundaries of the region
in which the absolute values of all three scattering lengths 
are greater than $2~\ell_{\rm vdW}$.
}
\label{fig:scatteringlengthslo}
\end{figure}

In Fig.~\ref{fig:scatteringlengthslo}, the three scattering lengths 
$a_{12}$, $a_{23}$, and $a_{13}$ are shown as 
functions of the magnetic field 
in the low-energy region from 0 to 600~G \cite{Julienne}.
Throughout most of this region, the smallest scattering length 
is $a_{12}$.  It satisfies $|a_{12}| > 2~\ell_{\rm vdW}$ in the interval
$122~{\rm G} < B < 485~{\rm G}$ and achieves its largest value 
$- 290~a_0 = -4.6~\ell_{\rm vdW}$ near 320~G.  This interval therefore contains a
universal region in which all three scattering lengths are 
negative and relatively large.
The zero-range approximation should be quantitatively useful 
in the middle of this interval,
but it becomes increasingly questionable as one approaches the edges.

\subsection{Three-body Recombination}

\begin{figure}[t]
\centerline{\includegraphics*[scale=0.7,angle=0,clip=true]{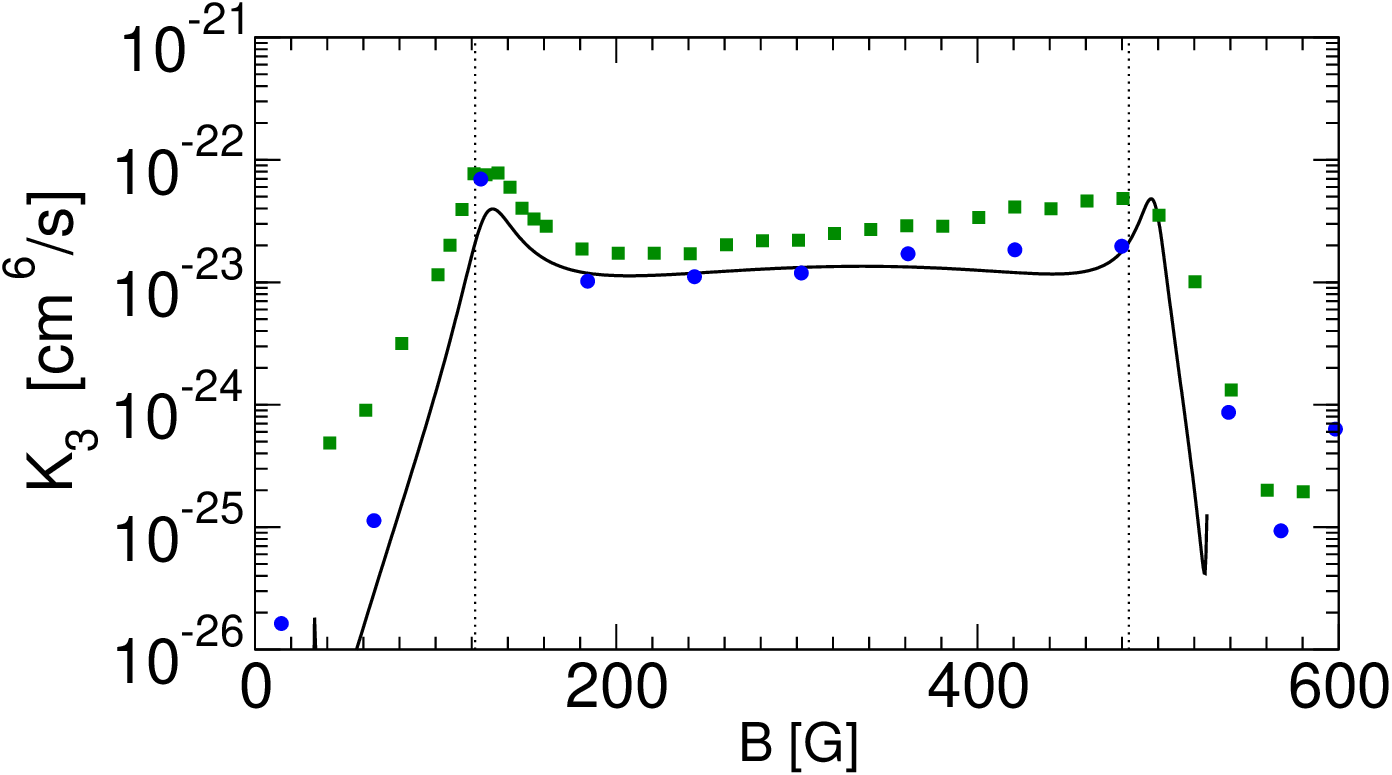}}
\vspace*{0.0cm}
\caption{(Color online) 
Three-body recombination rate constant $K_3$ 
as a function of the magnetic field $B$ from 0 to 600~G.
The solid squares and dots are data points from Refs.~\cite{Ottenstein:2008}
and \cite{Huckans:2008fq}, respectively.
The curve is the absolutely normalized result for $K_3^{\rm deep}$ with 
$\kappa_* = 76.8~a_0^{-1}$ and $\eta_* = 0.11$.
The two vertical dotted lines mark the boundaries of the region
in which the absolute values of all three scattering lengths 
are greater than $2~\ell_{\rm vdW}$.
}
\label{fig:Klow}
\end{figure}

The first measurements of the 3-body recombination rate $K_3$ for the three
lowest hyperfine spin states of $^6$Li atoms were carried out by Ottenstein et
al.~\cite{Ottenstein:2008} and by Huckans et al.~\cite{Huckans:2008fq}.  Their
results for magnetic field in the region from 0 to 600~G are shown in
Fig.~\ref{fig:Klow}. The results for $K_3$ in Ref.~\cite{Ottenstein:2008} from
measurements of the loss rates of the three individual spin states have been
averaged to get a single value of $K_3$ at each value of $B$.  Both groups
observed dramatic variations in $K_3$ with $B$, including a narrow loss
feature near 130~G and a broader loss feature near 500~G.

The narrow loss feature and the broad loss feature 
in the measurements of Refs.~\cite{Ottenstein:2008,Huckans:2008fq}
both appear near the boundaries of the
region in which all three scattering lengths satisfy 
$|a_{ij}| > 2 \ell_{\rm  vdW}$.  The effects of the finite range of the 
interaction  may be significant near the boundaries of this region.
In Ref.~\cite{Braaten:2008wd}, we fit the data for $K_3$ in this
region by calculating the 3-body recombination rate $K_3^{\rm deep}$ in
Eq.~(\ref{K3total-A}), using the magnetic field dependence of the three
scattering lengths shown in Fig.~\ref{fig:scatteringlengthslo} and treating
$\Lambda$ and $\eta_*$ as fitting parameters.  Since the systematic error in
the normalization of $K_3$ was estimated to be $90\%$ in
Ref.~\cite{Ottenstein:2008} and $70\%$ in Ref.~\cite{Huckans:2008fq}, we 
only fit the shape of the data and not its normalization.  
A 3-parameter fit to the
data from Ref.~\cite{Ottenstein:2008} in the region 
$122~{\rm G} < B < 485~{\rm G}$
with an adjustable normalization factor determines the 3-body parameters
$\Lambda = 436~a_0^{-1}$ and $\eta_* = 0.11$.  
This value of the cutoff is equivalent to $\kappa_* = 76.8~a_0^{-1}$.
These parameters $\kappa_*$ and $\eta_*$ determine
the normalization of $K_3^{\rm deep}$, so the theoretical curve 
in Fig.~\ref{fig:Klow} is absolutely normalized.
The fit to the shape of the narrow loss feature is excellent.  The
normalization is also correct to within the systematic error in the data.
However the fit predicts that $K_3$ should be almost constant in the middle of
the low-field region and that there should be another narrow loss feature at
its upper end near 500~G.  These predictions are not consistent with the data
in Fig.~\ref {fig:Klow}, which increases monotonically in the middle of the
low-field region and has a broad loss feature near the upper end of this
region.

Similar results for the 3-body recombination rate of $^6$Li atoms were
obtained subsequently by two other groups \cite{NU:2009,Schmidt:2008fz}.
Naidon and Ueda used hyperspherical methods to calculate the recombination
rate \cite{NU:2009}.  For their 3-body parameters, they used the real and
imaginary parts of the logarithmic derivative of the hyperradial wavefunction.
Schmidt, Floerchinger, and Wetterich used functional renormalization methods
to calculate the recombination rate \cite{Schmidt:2008fz}.  For their 3-body
parameters, they used the real and imaginary parts of the detuning energy of a
triatomic molecule.  The results of both groups for $K_3$ as a function of the
magnetic field are similar to our results in Fig.~\ref{fig:Klow}.  One
difference is that in Refs.~\cite{NU:2009,Schmidt:2008fz} the recombination
rate was calculated only up to an overall normalization constant that was
determined by fitting the data.  In our calculation, the absolute
normalization is determined by the 3-body parameters $\kappa_*$ and $\eta_*$.
\begin{figure}[t]
\centerline{\includegraphics*[scale=0.7,angle=0,clip=true]{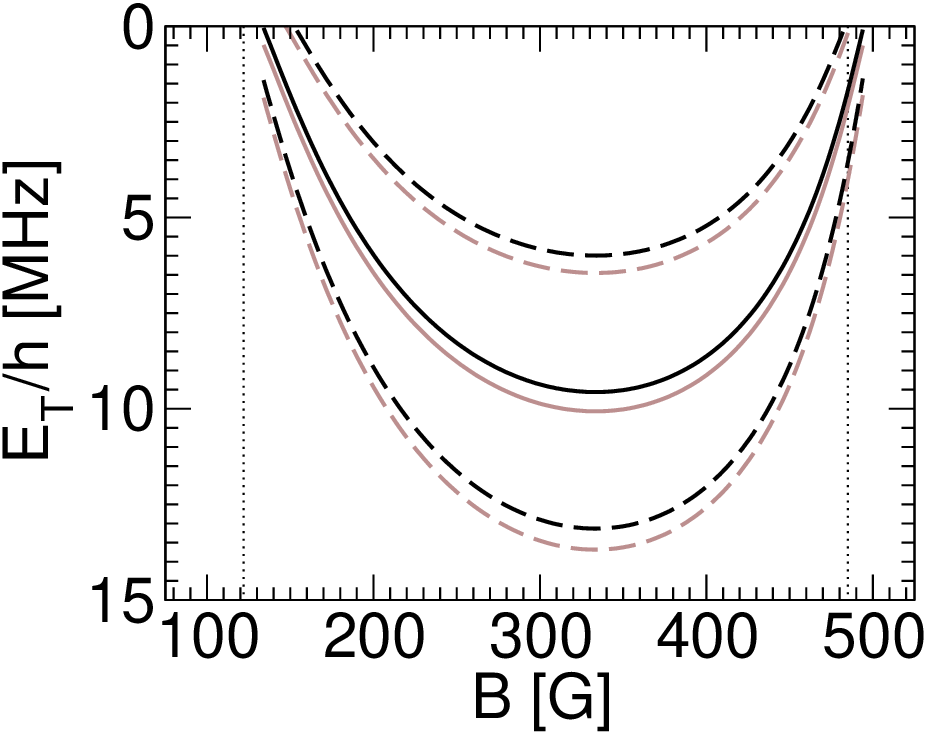}}
\vspace*{0.0cm}
\caption{(Color online) Energy 
  of the Efimov trimer as a function of the magnetic field $B$ 
  in the low-field region.  
  The binding frequency $E_T^{(1)}/(2 \pi \hbar)$ (dark solid line)
  and the frequency $\Gamma_T^{(1)}/(2 \pi\hbar)$ associated with the width
  (difference between dark dashed lines) were obtained from the 
  complex energy eigenvalue calculated using the parameters 
   $\kappa_* = 76.8~a_0^{-1}$ and $\eta_* = 0.11$.
   Also shown for comparison are the small-$\eta_*$ approximation 
   in  Eqs.~(\ref{E-approx}) and (\ref{Gamma-approx}) for the 
   binding frequency (light solid line)
  and for the frequency associated with the width 
  (difference between light dashed lines).
The two vertical dotted lines mark the boundaries of the region
in which the absolute values of all three scattering lengths 
are greater than $2~\ell_{\rm vdW}$.
}
\label{fig:energieslo}
\end{figure}

In Ref.~\cite{Wenz:2009}, Wenz et al.\ provided an explanation for the loss
feature near 500~G being much broader than predicted in
Refs.~\cite{Braaten:2008wd,NU:2009,Schmidt:2008fz}.  They pointed out that
there are deep dimers whose binding energies vary significantly over the
low-field region. These dimers are those responsible for the Feshbach
resonances 690~G, 811~G, and 834~G. In the high-field region, they are shallow
dimers, but they become deep dimers in the low-field region.  Their binding
energies change over the low-field universal region by as much as a factor of
6.  Wenz et al.\ assumed that contributions to the 3-body parameter 
$\eta_*$ scale like the inverse of the binding energy of the deep dimer.  
They used the
coefficient in this scaling relation as a fitting parameter along with the
3-body parameter equivalent to $\kappa_*$.  They obtained an excellent fit to
$K_3$ over the entire low-field region, including the narrow loss feature, the
broad loss feature, and the monotonic rise in between.  Their assumption for
the scaling of $\eta_*$ with the binding energy of the deep dimer can be
justified by an explicit calculation in a two-channel model
\cite{Braaten-Kang}.

In Ref.~\cite{Wenz:2009}, Wenz et al.\ proposed a simple analytic 
approximation for the 3-body recombination rate in regions 
where all 3 scattering lengths are negative.
Their approximation is the analytic result for equal scattering lengths 
in Eq.~(\ref{K3-equal}), with $a$ replaced by an effective scattering length 
$a_m$ given by
\begin{equation}
a_m = - \left[ (a_1^2 a_2^2 + a_2^2 a_3^2 + a_3^2 a_1^2)/3 \right]^{1/4}.
\label{am}
\end{equation}
Another possible choice for an effective scattering length 
is the geometric mean of the three scattering lengths:
\begin{equation}
a_g = - | a_1 a_2 a_3|^{1/3}.
\label{ag}
\end{equation}
We can use our universal results to test the accuracy of that 
approximation. We find that the analytic 
result in Eq.~(\ref{K3-equal}) with $a$ replaced by 
the geometric mean $a_g$ in Eq.~(\ref{ag}) is a significantly 
more accurate approximation.

\subsection{Efimov Trimers}
\label{sec:EfTrilo}

Naidon and Ueda \cite{NU:2009}
and Schmidt, Floerchinger, and Wetterich \cite{Schmidt:2008fz}
calculated the binding energy of the Efimov trimer 
that is responsible for the loss features in Fig.~\ref{fig:Klow}.
Its binding energy goes to 0 near both the observed narrow loss feature 
near 130~G and the predicted narrow loss feature near 500~G.  This
demonstrates explicitly that both loss features
arise from the same Efimov trimer crossing the 3-atom threshold.

We calculate the spectrum of Efimov trimers by solving for the complex energy
eigenvalues of the homogeneous STM equations in Eq.~(\ref{STMeqB}).  The
energies and widths of the trimers are obtained by expressing the complex
energies in the form $-E_T^{(n)} - i \Gamma_T^{(n)}/2$.  We choose to label
the Efimov trimer responsible for the narrow loss feature near 135~G by the
integer $n=1$. Our predictions for the binding energy $E_T^{(1)}$ of the
shallowest Efimov state are shown in
Fig.~\ref{fig:energieslo} as a function of the magnetic field.  The binding
frequency $E_T^{(1)}/(2 \pi \hbar)$ increases from 0 at $134$~G to a maximum
of $9.56$~MHz at $334$~G and then decreases to 0 at $494$~G.  Our maximum
binding frequency agrees well with the maximum frequencies of about 10~MHz and
about 11~MHz obtained in Refs.~\cite{NU:2009} and \cite{Schmidt:2008fz},
respectively.  The binding frequency $E_T^{(0)}/(2 \pi \hbar)$ associated with
the next deeper Efimov trimer is predicted to change gradually from $12.1$~GHz
at $134$~G to $12.5$~GHz at $334$~G and then to $12.1$~GHz at $494$~G.  Since
this binding energy is much larger than the van der Waals frequency
$\nu_{\textrm{vdW}} = 154~{\textrm{MHz}}$, this Efimov trimer and all the
deeper ones predicted by the STM equations are artifacts of the zero-range
approximation. 
Our value of $\kappa_*$ can be interpreted as the binding wavenumber
$(m E_T^{(-2)}/\hbar^2)^{1/2}$ of the fictitious Efimov trimer labelled
by $n=-2$.  It corresponds to the choice $n_*=-2$ in Eq. (\ref{Efimov}).

We note that the finite range corrections to the binding energies of 
the deeper trimers can be substantial if their energy becomes comparable to
the van der Waals energy even if all scattering lengths are large.
The leading corrections to the binding energy for the deep trimers
are of order $\ell_{vdW}(m E_T^{(n)}/\hbar^2)^{1/2}$. For the trimer
labeled $n=1$, these corrections are 25\% for the largest
value of the binding energy.

In Ref.~\cite{NU:2009}, Naidon and Ueda also calculated the width of the
Efimov trimer.  In Fig.~\ref{fig:energieslo}, our results for the width
$\Gamma_T^{(1)}$ are illustrated by plotting the frequencies associated with
the energies $E_T^{(1)} \pm \Gamma_T^{(1)}/2$ as functions of the magnetic
field.  The frequency $\Gamma_T^{(1)}/(2 \pi \hbar)$ increases from $2.73$~MHz
at $134$~G to a maximum of $7.14$~MHz at $332$~G and then decreases to
$2.56$~MHz at $494$~G. Our maximum width is more than twice as large as the
maximum width of about (3~MHz)$\times 2 \pi \hbar$ obtained in
Ref.~\cite{NU:2009}. Since our result for $\Gamma_T^{(1)}$ is 
always comparable to or larger than $E_T^{(1)}$, the interpretations
of $E_T^{(1)}$ and $\Gamma_T^{(1)}$ as the binding energy and width of the
trimer should be viewed with caution.

Since $\eta_* = 0.11$ is relatively small, we can also use the small-$\eta_*$
approximations to the binding energy and the width given in
Eqs.~(\ref{Gamma-approx}) and (\ref{E-approx}).  These approximations, which
are calculated using $\kappa_* = 76.8~a_0^{-1}$, are shown in
Fig.~\ref{fig:energieslo} for comparison.  The
maximum binding frequency $E_T^{(1)}/(2 \pi \hbar)$ is $10.1$~MHz at $334$~G,
which is larger than the result from the complex energy by about 6\%.  
The frequency $\Gamma_T^{(1)}/(2 \pi \hbar)$ increases 
from $2.76$~MHz at $134$~G
to a maximum of $7.23$~MHz at $332$~G and then decreases to $2.59$~MHz at
$494$~G. These values are larger than the results from the complex energy 
by only about 1\%.

\section{$\bm{^6}$Li Atoms: High-field Universal Region}
\label{sec:hifield}

\begin{figure}[t]
\centerline{\includegraphics*[scale=0.7,angle=0,clip=true]{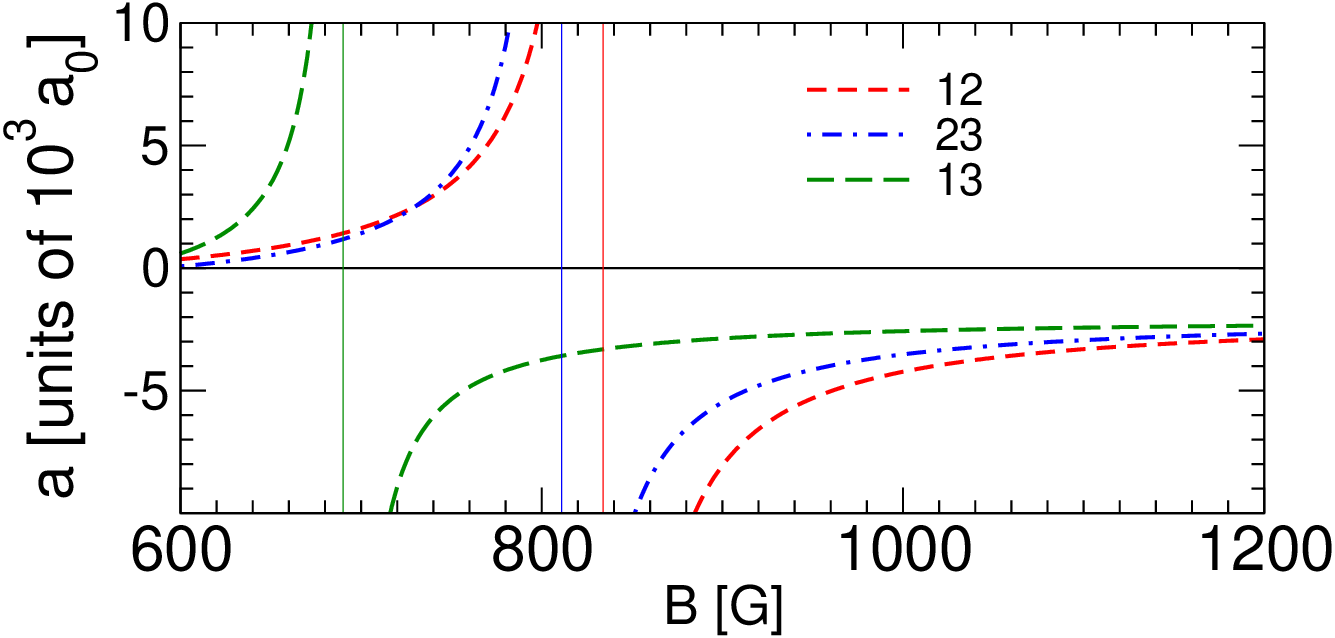}}
\vspace*{0.0cm}
\caption{(Color online) 
The scattering lengths in units of $10^3 a_0$ 
for the three lowest hyperfine states 
of $^6$Li as functions of the magnetic field $B$ 
from 600~G to 1200~G \cite{Julienne}.
The three vertical lines mark the positions of the 
Feshbach resonances. 
}
\label{fig:scatteringlengthshi}
\end{figure}

In this section, we apply our formalism to the lowest three hyperfine spin
states of $^6$Li atoms in the region of high magnetic field from 600 to
1200~G. In Fig.~\ref{fig:scatteringlengthshi}, the three scattering lengths
$a_{12}$, $a_{23}$, and $a_{13}$ are shown as functions of the magnetic field
\cite{Julienne}.  This region includes the Feshbach resonances in $a_{12}$,
$a_{23}$, and $a_{13}$ near 834~G, 811~G, and 690~G, respectively.  Beyond
these Feshbach resonances, all three scattering lengths approach the triplet
scattering length $-2140~a_0$.  For $B > 637$~G, the absolute values of all
three scattering lengths are larger than $2140~a_0 \approx 34~\ell_{\rm vdW}$.
An estimate of the lower boundary of this universal region is 608~G, where the
smallest scattering length is $a_{13} = 125~a_0 \approx 2 \ell_{\rm vdW}$.
The zero-range approximation should be very accurate throughout most of this
high-field region.  The physics in this universal region is rich, with the
Feshbach resonances marking the boundaries between regions in which 3, 2, 1,
or 0 of the three scattering lengths are positive.

\subsection{Measurements of Three-body Recombination}
\label{sec:3BRhi}

In Ref.~\cite{Huckans:2008fq}, Huckans et al.\ presented measurements 
of the 3-body recombination rate at 6 values of the magnetic field 
in the range 600~G to 1000~G.  The low-temperature limit of the 
recombination rate could not be measured, because the temperature was not low
enough and because of heating associated with the Feshbach resonance.
In Ref.~\cite{Braaten:2008wd}, we showed that a naive fit of the last 
two data points at 894~G and 953~G using our zero-temperature
calculations indicated an Efimov resonance near 1200~G.
We suggested that it might be worthwhile to search the high-field 
region for an Efimov resonance.

\begin{figure}[t]
\centerline{\includegraphics*[scale=0.7,angle=0,clip=true]{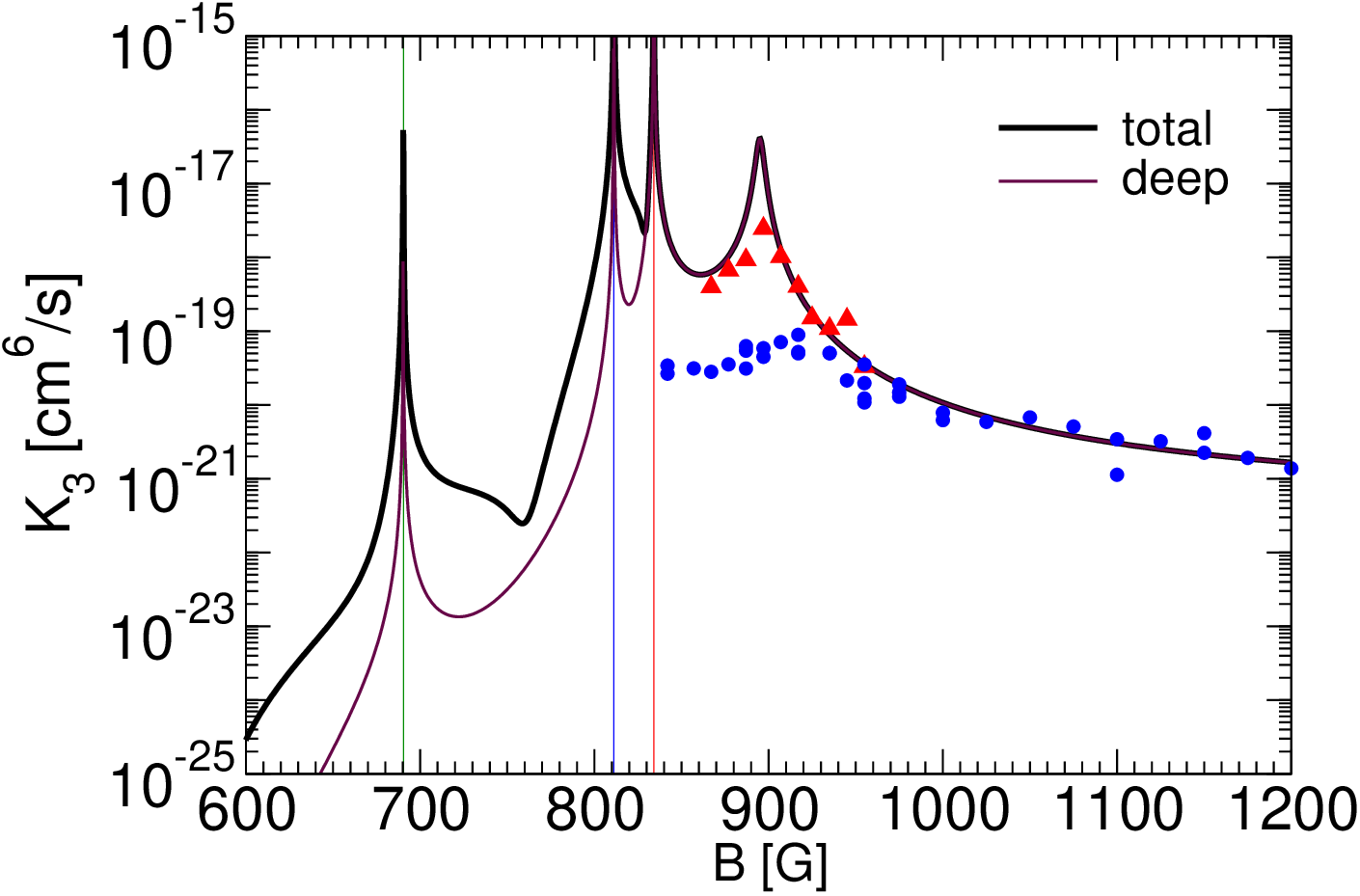}}
\vspace*{0.0cm}
\caption{(Color online) Three-body recombination rate constant $K_3$ as a
  function of the magnetic field $B$ from 600~G to 1200~G.  The solid dots 
  and triangles are data points from Ref.~\cite{Huckans:2008fq} 
  at temperatures less than 180~nK and less than 30~nK, respectively.
  The curve between 834~G and 1200~G is a fit to that data.  
  In the region from 600~G and 834~G, the curves are our predictions 
  for the total 3-body recombination rate (thick line) 
  and the contribution from recombination into deep dimers (thin line)
  for $\kappa_* = 80.7~a_0^{-1}$ and $\eta_* = 0.016$.
  The three vertical lines mark the positions of the 
  Feshbach resonances. 
}
\label{fig:Khigh}
\end{figure}

A narrow 3-atom loss resonance in the high-field region 
near 895~G was recently 
discovered by Williams et al.~\cite{Williams:2009} and by Jochim and
coworkers~\cite{Jochim}.  Williams et al.\ measured the 3-body recombination
rate at magnetic fields from 842~G up to 1500~G at temperatures lower 
than 180~nK and from 834~G up to 955~G at temperatures lower than 30~nK. 
Their data is shown in Fig.~\ref{fig:Khigh}. 
By fitting their measurements using our formalism, they obtained the
3-body parameters $\kappa_* = (6.9 \pm 0.2)\times 10^{-3}~a_0^{-1}$ and
$\eta_* = 0.016^{+0.006}_{-0.010}$~\cite{Williams:2009}.  Since $\kappa_*$ is
only defined up to multiplication by integer powers of 
$\lambda_0 \approx 22.7$, an equivalent value is 
$\kappa_* = 80.7 \pm 2.3~a_0^{-1}$.  Their fit
for the central values of $\kappa_*$ and $\eta_*$ is shown in
Fig.~\ref{fig:Khigh} as the thick solid line between 834~G and 1200~G.
The fitted value for the position of the 3-atom loss resonance
is $895^{+4}_{-5}$~G. The peak value of the 3-body recombination
rate at zero temperature is predicted to be
$(4.1^{+8.5}_{-1.5}) \times 10^{-17}$~cm$^6$/s.
 
It is interesting to compare the values of the 3-body parameters
$\kappa_*$ and $\eta_*$ in the high-field region
with those in the low-field region.
The values in the low-field region obtained by fitting the measurements
of the narrow 3-body
recombination loss feature by Ottenstein et al.\  \cite{Ottenstein:2008}
were $\kappa_* \approx 76.8~a_0^{-1}$ and $\eta_* \approx 0.11$.
It is difficult to quantify the errors in these parameters,
because the narrow loss feature
is at the edge of the universal region where range corrections may be
significant.  The values of $\kappa_*$ in the high-field region
and the low-field region are consistent within errors.
This could be just a coincidence, but
it suggests that the phase in the 3-body wavefunction
at short distances that controls Efimov physics
is insensitive to the Feshbach resonances that
change the scattering lengths.
The value of $\eta_*$ in the high-field region is about an
order-of-magnitude smaller than in the low-field region.
This demonstrates that the 3-body parameters need not be equal
in the two universal regions.  Wenz et al.\ proposed
a mechanism for variations in $\eta_*$, namely
significant changes in the binding energies of deep dimers
with the magnetic field \cite{Wenz:2009}.
This mechanism might also explain the order-of-magnitude
difference in $\eta_*$ between the two regions.

Using the values of $\kappa_*$ and $\eta_*$
determined by O'Hara et al.~\cite{Williams:2009},
we can calculate the universal predictions for 
other aspects of Efimov physics for $^6$Li atoms 
in the high-field universal region.
The error bars on the values of $\kappa_*$ and $\eta_*$ 
can be used to give error bars on the predictions.
In the next subsections, we give universal predictions 
for the binding energies and the widths of the Efimov trimers
and for the 3-body recombination rate.

\subsection{Efimov Trimers}
\label{sec:EfTrihi}

\begin{figure}[t]
\centerline{\includegraphics*[scale=0.7,angle=0,clip=true]{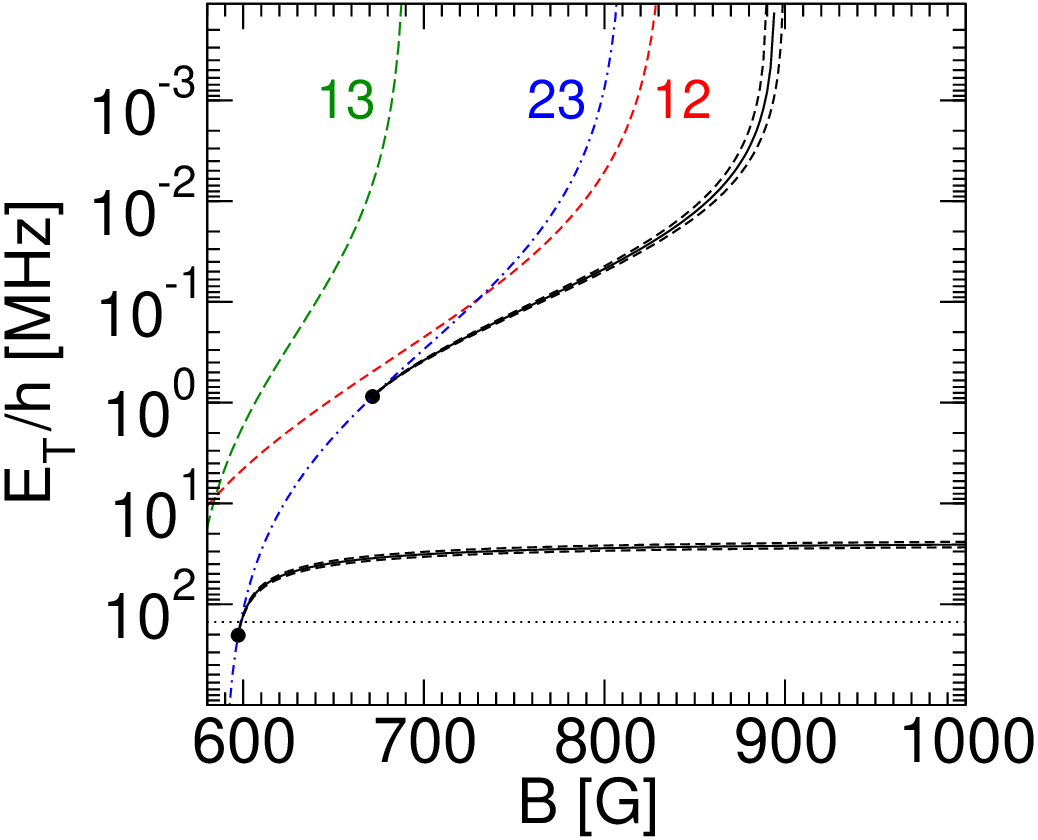}}
\vspace*{0.0cm}
\caption{(Color online) 
Energies of the Efimov trimers as functions 
of the magnetic field $B$ in the high-field region.
The solid curves are the predicted binding frequencies 
$E_T^{(n)}/(2 \pi \hbar)$ 
for $\kappa_* = 80.7~a_0^{-1}$ and $\eta_* = 0.016$.
The dashed curves are the upper and lower error bars obtained 
by varying $\kappa_*$.
The curves labelled 12, 23, and 13 are the atom-dimer thresholds.
The dots indicate the points where the trimers disappear 
through the 1+(23) atom-dimer threshold.
The horizontal dotted line is the van der Waals frequency 154~MHz.  }
\label{fig:energieshi}
\end{figure}

We calculate the binding energies and the widths of the Efimov trimers by
solving for the complex energy eigenvalues of the homogeneous STM equations in
Eq.~(\ref{STMeqB}).  We choose to label the Efimov trimer responsible for the
narrow loss feature near 895~G by the integer $n=1$.  In
Fig.~\ref{fig:energieshi}, the predicted binding energies $E_T^{(0)}$ and
$E_T^{(1)}$ of the two shallowest Efimov trimers are shown as functions of the
magnetic field.  The Efimov trimer responsible for the narrow loss feature has
a binding energy $E_T^{(1)}$ that vanishes at $895^{+4}_{-5}$~G.  As the
magnetic field decreases, $E_T^{(1)}$ increases monotonically until the
critical value $B_* = 672 \pm 2$~G where the Efimov trimer disappears through
the 1+(23) atom-dimer threshold.  When it disappears, its binding energy
$E_T^{(1)}$ relative to the 3-atom threshold is $871 ^{+43}_{-68}$~kHz, which
is also the binding energy $\hbar^2/(m a_{23}^2)$ of the (23) dimer.  The next
deeper Efimov trimer has a binding frequency $E_T^{(0)}/(2 \pi \hbar)$ that
increases monotonically from $26.4^{+1.7}_{-1.6}$~MHz at $895$~G to
$34.9^{+1.9}_{-1.9}$~MHz at $B_*$. 
Our solutions to the STM equations predict that it 
crosses the 1+(23) atom-dimer threshold at 597~G,
where its binding frequency relative to the 3-atom threshold is $203$~MHz.
This frequency is larger than the van der Waals frequency 154~MHz.
Moreover the smallest scattering length at this point is only $54~a_0$,
which is smaller than the van der Waals length $65~a_0$.  
The zero-range predictions are not expected to be accurate 
for such large energies and for such small scattering lengths.  
All the
deeper Efimov trimers have binding energies that are much larger than the van
der Waals energy and are therefore artifacts of the zero-range approximation.
Our value of $\kappa_*$ can be interpreted as the binding wavenumber
$(m E_T^{(-2)}/\hbar^2)^{1/2}$ of the fictitious Efimov trimer labelled
by $n=-2$.  It corresponds to the choice $n_*=-2$ in Eq. (\ref{Efimov}).


As discussed in Sec.~\ref{sec:EfTrilo},
the leading range corrections to the binding energy for the deep trimers
are estimated to be of order $\ell_{vdW}(m E_T^{(n)}/\hbar^2)^{1/2}$.
For the trimer labeled $n=1$, these corrections are 8\% where it crosses
the 1+(23) atom-dimer threshold and become significantly smaller at
larger values of the magnetic field. For the trimer labelled $n=0$
these corrections are about 40\% at $B_*$ and larger values of the
magnetic field.

\begin{figure}[t]
\centerline{\includegraphics*[scale=0.7,angle=0,clip=true]{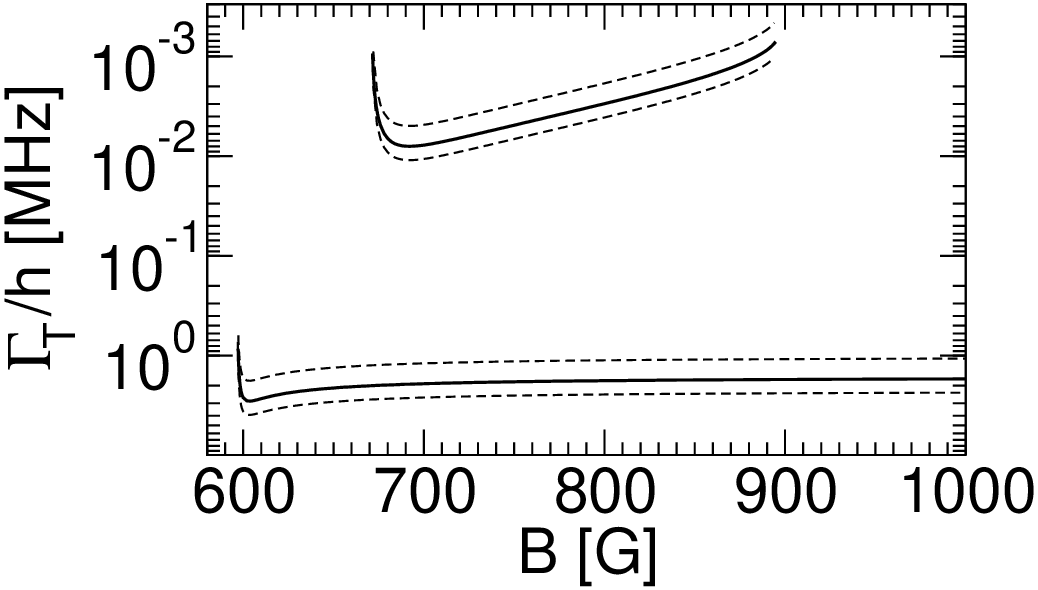}}
\vspace*{0.0cm}
\caption{
Widths of the Efimov trimers as functions of the magnetic field $B$  
in the high-field region.   The solid curves are the 
frequencies $\Gamma_T^{(n)}/(2 \pi \hbar)$ associated with the widths
for $\kappa_* = 80.7~a_0^{-1}$ and $\eta_* = 0.016$.
The dashed curves are the upper and lower error bars obtained 
by varying $\eta_*$.
}
\label{fig:widthshi}
\end{figure}

In Fig.~\ref{fig:widthshi}, the predicted widths $\Gamma_T^{(n)}$ of the Efimov
trimers are shown as functions of the magnetic field.  The frequency
$\Gamma_T^{(1)}/(2 \pi \hbar)$ associated with 
the width of the shallower trimer increases from
$0.706$~kHz at $895$~G to a maximum of $7.98$~kHz at $692$~G and then
decreases to $0.940$~kHz at $B_*$.  The frequency $\Gamma_T^{(0)}/(2 \pi \hbar)$
associated with the width of the deeper trimer
increases from $1.74$~MHz at $895$~G to $2.00$~MHz
at $B_*$.

\begin{figure}[t]
\centerline{\includegraphics*[scale=0.7,angle=0,clip=true]{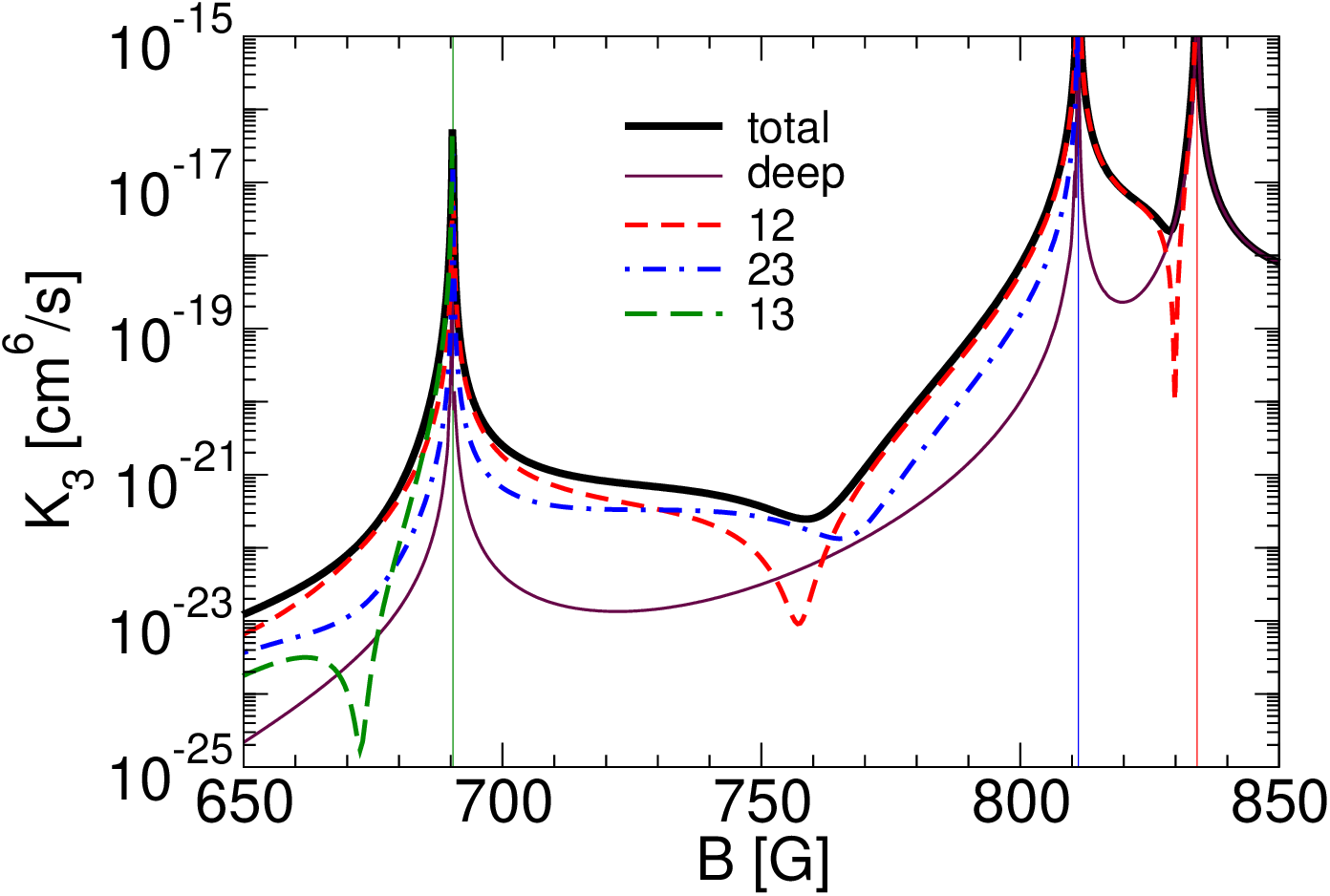}}
\vspace*{0.0cm}
\caption{(Color online) 
Three-body recombination rate constant $K_3$ 
as a function of the magnetic field $B$ from 650~G to 850~G.
The curves are our predictions 
for the total 3-body recombination rate (thick solid line), 
the contribution from recombination into deep dimers (thin solid line),
and the contributions from recombination into 
the (12) dimer (short-dashed line), the (23) dimer (dash-dotted line),
and the (13) dimer (long-dashed line)
for $\kappa_* = 80.7~a_0^{-1}$ and $\eta_* = 0.016$.}
\label{fig:Khighlo}
\end{figure}

\subsection{Predictions for Three-body Recombination}
\label{sec:3BRhipre}

Our predictions for the 3-body recombination rate in the region 600~G to 834~G,
where one or more of the scattering lengths are positive,
are shown in Fig.~\ref{fig:Khigh}.  The upper curve is the total
recombination rate $K_3$, while the lower one is the contribution $K_3^{\rm
  deep}$ from deep dimers.  Besides the Efimov resonance at 895~G, the only
other peaks are at the Feshbach resonances at 690~G, 811~G, and 834~G.  They
arise simply from the scaling of $K_3$ as $a^4$ up to logarithms, where $a$ is
some appropriate mean of the scattering lengths $a_{12}$, $a_{23}$, and
$a_{13}$. The total recombination rate $K_3$ is predicted to have three local
minima. There is a broad local minimum at $759 \pm 1$~G, where the minimum
recombination rate is $(2.5^{+0.5}_{-0.5}) \times 10^{-22}$~cm$^6$/s.  There
is a narrow local minimum at $829 \pm 1$~G, where the minimum recombination
rate is $(2.2^{+0.9}_{-1.3}) \times 10^{-18}$~cm$^6$/s.  Finally there is a
local minimum at $861 \pm 2$~G between the last Feshbach resonance and the
3-atom Efimov resonance, where the minimum recombination rate is
$(5.8^{+3.8}_{-4.0}) \times 10^{-19}$~cm$^6$/s.

In Fig.~\ref{fig:Khighlo}, we show the predictions for the 
3-body recombination rate in the region from 650~G to 850~G in more detail.  
In addition to $K_3$ and $K_3^{\rm deep}$, we show the 
contributions from recombination into the (12), (23), and (13) dimers.
This figure reveals that the local minima at 759~G and 829~G
are associated with interference in the recombination into shallow dimers.
The narrow minimum at 829~G arises from the combination of a rapidly 
increasing rate into deep dimers and a rapidly decreasing rate into 
(12) dimers, which has its minimum at 830~G.
The broad minimum at 759~G arises from minima in the rates into 
(12) and (23) dimers at 757~G and 765~G, respectively.
There are also minima in the rates into the (13) dimer at 672~G
and the (23) dimer at 600~G,
but their effects are not visible in the total recombination rate.
We have verified that these minima in the recombination rates into
individual shallow dimers arise from interference effects by showing that they
become zeroes as $\eta_*$ is decreased to 0.  Thus these minima
are the result of destructive interference between two recombination pathways.
In the case of three equal positive scattering lengths, a similar 
interference effect
is evident in the analytic expression for the 3-body recombination rate into
shallow dimers in Eq.~(\ref{alpha-analytic:eta}).

\subsection{Atom-Dimer Resonance}
\label{sec:ADRes}

An atom-dimer loss resonance can appear at a value of the scattering length 
for which an Efimov trimer crosses the atom-dimer threshold.
From the Efimov trimer spectrum in  Fig.~\ref{fig:energieshi}, 
one can see that atom-dimer resonances are predicted at the two values 
of the magnetic field where the Efimov trimers cross the 
1+(23) atom-dimer threshold.  The atom-dimer resonance associated with
the shallower of the two Efimov trimers is predicted to occur 
at $B_* = 672 \pm 2$~G.
The universal predictions for the dimer relaxation rate near this resonance 
can be calculated by solving appropriate sets of coupled STM equations. 
The atom-dimer resonance associated with the deeper of the 
two Efimov trimers is predicted to occur at $B_*' = 597$~G.
This resonance occurs slightly outside the universal region,
so universal predictions for the position of the resonance 
and for the dimer relaxation rate are not expected to be accurate.

If we restrict our attention to magnetic fields very near the atom-dimer
resonance at $B_*$, we can get an approximation to the universal predictions
for the dimer relaxation rate $\beta_{1(23)}$ without actually solving the STM
equations. We take advantage of the fact that the atom-dimer scattering length
$a_{1(23)}$ diverges at $B_*$.  The 3-atom problem in this region therefore
reduces to a universal 2-body problem for the atom and the (12) dimer.  The
universal properties are determined by the large scattering length
$a_{1(23)}$.  For example, in the region $B > B_*$ but very close to $B_*$,
the binding energy of the Efimov trimer is well approximated by the sum of the
binding energy of the (23) dimer and a universal term determined by the
atom-dimer scattering length:
\begin{equation}
E^{(1)}_T \approx - \left( \frac{\hbar^2}{m a_{23}^2} 
                       + \frac{3\hbar^2}{4 m a_{1(23)}^2} \right).
\label{ET-aAD}
\end{equation}
This should be a good approximation as long as the second term 
is much smaller than the first term.
In the region near $B_*$, the atom-dimer scattering length can be 
approximated by an expression that has the same form as the analytic result  
for identical bosons but with different numerical coefficients: 
\begin{equation}
a_{1(23)} \approx \left( C_1 \cot[ s_0 \ln(a_{23}/a_*) + i \eta_*]
                         + C_2 \right) a_{23} ,
\label{aAD-a23}
\end{equation}
where $a_* = a_{23}(B_*)$.  The coefficients $C_1$ and $C_2$ can in
principle depend on ratios of the scattering lengths, but $a_{13}$ is an order
of magnitude larger so it essentially decouples and $a_{13}$ and $a_{23}$ are
both increasing with $B$ so their ratio does not change rapidly near $B_*$.
Thus we can treat $C_1$ and $C_2$ as numerical constants.  These
constants can be determined by using Eq.~(\ref{ET-aAD}) to extract
$a_{1(23)}$ from the results for the real part of the trimer binding energy
shown in Fig.~\ref{fig:energieshi} and then fitting those results to the
expression for the atom-dimer scattering length in Eq.~(\ref{aAD-a23}). The
resulting constants are $C_1 = 0.67$ and $C_2 = 0.65$.

\begin{figure}[t]
\centerline{\includegraphics*[scale=0.7,angle=0,clip=true]{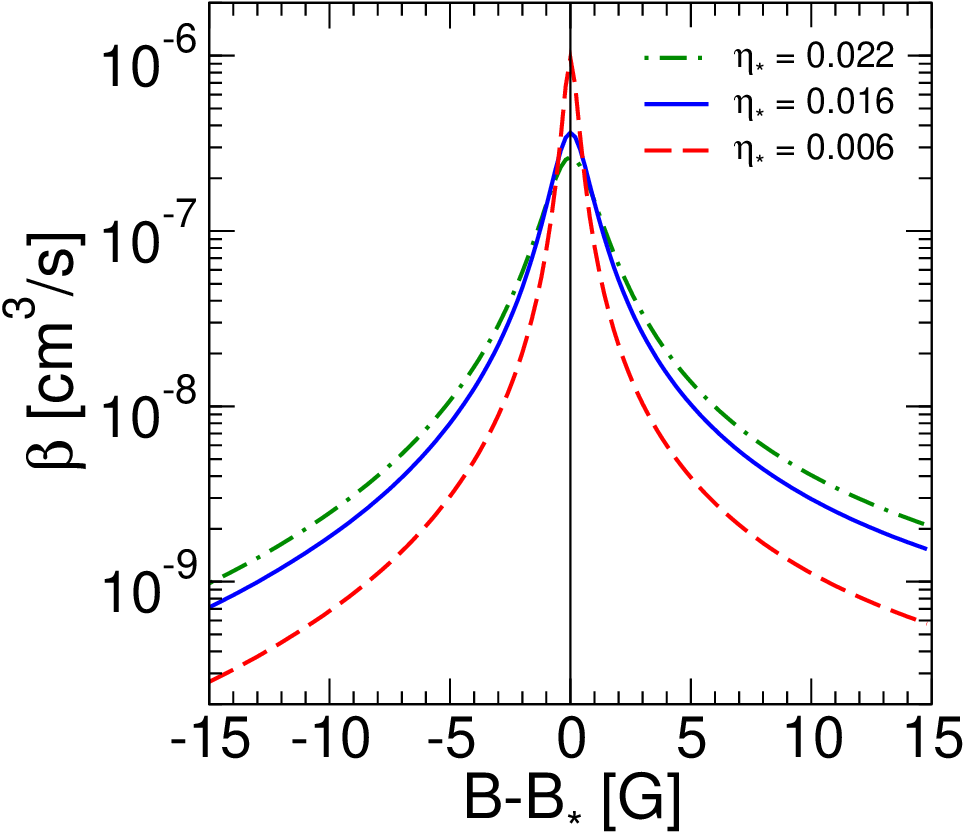}}
\vspace*{0.0cm}
\caption{(Color online) 
Dimer relaxation rate constant $\beta_{1(23)}$  
for (23) dimers and atoms of type 1
as a function of the magnetic field $B$. 
The magnetic field is measured relative to the position $B_*$ 
of the atom-dimer resonance.
The curves are predictions using the approximation in 
Eq.~(\ref{beta-a23}) for three values of $\eta_*$:
0.006, 0.016, and 0.022.}
\label{fig:beta}
\end{figure}

With the constant $C_1$ in hand, 
we can obtain an approximation for the dimer relaxation 
rate constant $\beta_{1(23)}$ for $B$ near $B_*$ simply by inserting the 
approximation for $a_{1(23)}$ in Eq.~(\ref{aAD-a23}) 
into the optical theorem relation in
Eq.~(\ref{beta-opt}).  The resulting expression is 
\begin{equation}
\beta_{1(23)} \approx 
\frac{9.44 \, C_1 \sinh (2 \eta_*)}
     {\sin^2[ s_0 \ln(a_{23}/a_*)] + \sinh^2 \eta_*} \, 
\frac{\hbar a_{23}}{m}.
\label{beta-a23}
\end{equation}
If $B_* = 672$~G, the value of $a_*$ is $835~a_0$.
The only dependence on the magnetic field is through the $B$-dependence
of $a_{23}$.  This approximation for $\beta_{1(23)}$ is shown as a function of
$B-B_*$ in Fig.~\ref{fig:beta}. The maximum value of $\beta_{1(23)}$ 
at the peak of
the resonance is $4 \times 10^{-7}$~cm$^3$/s for $\eta_* = 0.016$.

Since the atom-dimer resonance associated with the deeper of the 
two Efimov trimers occurs slightly outside the universal region,
universal predictions for the dimer relaxation rate are not 
expected to be accurate.  They might however be useful 
for making order-of-magnitude estimates.
If we apply the analysis described above to the 
binding energy of the deeper Efimov trimer, we get the constants
$C_1 = 0.7$ and $C_2 = 0.3$.
The value of $a_*$ is approximately $54~a_0$.
If we use Eq.~(\ref{beta-a23}) to estimate the dimer relaxation rate, 
we obtain a maximum value of $\beta_{1(23)}$ at the peak of
the resonance of $2.4 \times 10^{-8}$~cm$^3$/s for $\eta_* = 0.016$.

\subsection{Many-body physics}
\label{sec:manybody}

One of the most important motivations for understanding few-body physics is
that there are aspects of many-body physics that are controlled by few-body
observables.  Our universal results for the few-body physics of $^6$Li atoms
with 3 spin states provides useful information about many-body systems of
$^6$Li atoms with sufficiently low temperature and sufficiently low number
densities.  If the atoms are in thermal equilibrium at temperature $T$, an
important scale is the thermal length $\lambda_T =(2 \pi m k_B
T/\hbar^2)^{-1/2}$.  If the atoms of type $i$ have number density $n_i$, another
important scale is the Fermi wavenumber $k_{Fi} =(3 \pi^2 n_i)^{1/3}$.  In our
calculation of the 3-body recombination rate, we set the
wavenumbers of the
incoming atoms to 0.  In order for our result to apply quantitatively to all
atoms in the system, it is necessary that $\lambda_T^{-1}$, $k_{F1}$,
$k_{F2}$, and $k_{F3}$ all be small compared to the wavenumber scale $(1/a_1^2
+ 1/a_2^2 + 1/a_3^2)^{1/2}$ set by the interactions.  Even if this condition
is not well satisfied, our result can be used to obtain order-of-magnitude
estimates of the relevant time scales.

There are several relevant time scales that we can extract from our few-body
calculations.  If the system contains Efimov trimers, an important time scale
is the lifetime $\hbar/\Gamma_T^{(n)}$ of the Efimov trimer, where
$\Gamma_T^{(n)}$ is its width.  The universal predictions for the widths of
the Efimov trimers are shown in Fig.~\ref{fig:widthshi}.  If the system can be
approximated by a gas of individual low-energy atoms, the time scale for
disappearance of a significant fraction of the atoms is set by the 3-body
recombination rate constant $K_3$.  From the rate equation in
Eq.~(\ref{dndt}), we see that the time scale for loss of a significant
fraction of the atoms of type $i$ is $(K_3 n_j n_k)^{-1}$, where $j$ and $k$
are the two complimentary spin states.  The universal predictions for the
3-body recombination rate are shown in Figs.~\ref{fig:Khigh} and
\ref{fig:Khighlo}.  Finally if the system contains low-energy dimers and
low-energy atoms in the complimentary spin state, the time scale for
disappearance of a significant fraction of the atoms or dimers is set by the
appropriate dimer relaxation rate constant.  From the rate equation in
Eq.~(\ref{beta-def}), we see that the time scales for loss of significant
fractions of (23) dimers and of atoms of type 1 are $(\beta_{1(23)}
n_{1})^{-1}$ and $(\beta_{1(23)} n_{(23)})^{-1}$, respectively.  The only
universal information we have about dimer relaxation rates is the
approximation for $\beta_{1(23)}$ in Eq.~(\ref{beta-a23}), which should be
accurate for magnetic fields within about 10~G of $B_* \approx 672$~G.  For
magnetic fields further from this critical value but still within the region
where $a_{23}$ is positive, one may be able to use the extrapolation of this
expression as an estimate for the dimer relaxation rate.

If the system contains a Bose-Einstein condensate of low-energy dimers 
as well as low-energy atoms in the complimentary spin state,
another important few-body observable is the atom-dimer scattering length
for that particular atom and dimer.  
The real part of the atom-dimer scattering length determines 
the mean-field energy of the atom in the dimer condensate.
The mean-field energy of an atom of type 1 in the (23) dimer condensate
with sufficiently low number density $n_{(23)}$
is $(3 \pi \hbar^2/m) \, {\rm Re} (a_{1(23)}) n_{(23)}$.
The approximation for $a_{1(23)}$ in Eq.~(\ref{aAD-a23})
should be accurate for magnetic fields within about 10~G 
of $B_* \approx 672$~G.  

We will illustrate the relevance of our universal result to 
many-body physics by applying them to two specific values of the 
scattering length that are interesting from a symmetry perspective.
If the pair scattering lengths $a_{ij}$ and $a_{ik}$ are equal, 
atoms of types $j$ and $k$ are related by an $SU(2)$ symmetry.
If all three scattering lengths are equal, the three spin states
are related by an $SU(3)$ symmetry.  There is an $SU(2)$ symmetry
point at 731~G, where the scattering lengths are 
$a_{12} = a_{23} \approx +2500~a_0$ and $a_{13} \approx  -7100~a_0$. 
An $SU(3)$ symmetry point can be approached by going to very 
high magnetic fields, where
all three scattering lengths approach the spin-triplet scattering length 
$-2140~a_0$ \cite{Bartenstein:2005}.

A quantum degenerate Fermi gas of $^6$Li atoms with approximately equal
populations of the three lowest hyperfine spin states at 1500~G has been
realized by Williams et al.\ \cite{Williams:2009}.  The scattering lengths are
$a_{12} \approx -2460~a_0$, $a_{23} \approx -2360~a_0$, and $a_{13} \approx
-2240~a_0$, so there is an approximate $SU(3)$ symmetry.  One candidate for a
metastable ground state is a state that can be approximated by filled Fermi
spheres for all three spin states, with Cooper pairing that breaks the $SU(3)$
symmetry down to a $U(1)$ subgroup. The lifetime for such a system is
determined by the 3-body recombination rate $K_3$, which is predicted to
be about $8 \times 10^{-22}$~cm$^6$/s at 1500~G.  Another
candidate for a metastable ground state is a filled Fermi sphere of Efimov
trimers.  The binding energy $E_T^{(0)}$ of the Efimov trimer is predicted to
be about 25~MHz$\times (2 \pi \hbar)$ at 1500~G. 
Its width $\Gamma_T^{(0)}$ is predicted to be 
about 1.7~MHz$\times (2 \pi \hbar)$, which corresponds to a lifetime
$\hbar/\Gamma_T^{(0)}$ of about $9 \times 10^{-8}$~s.

We now consider a many-body system at 731~G, where there is an $SU(2)$
symmetry relating the atoms of types 1 and 3.  For a state that can be
approximated by filled Fermi spheres for all three spin states, the lifetime
is determined by the the 3-body recombination rate $K_3$, which is predicted
to be about $7 \times 10^{-22}$~cm$^6$/s at 731~G. A better candidate for a
metastable ground state is a state containing a Bose-Einstein condensate of
dimers, which breaks the $SU(2)$ symmetry down to a $U(1)$ subgroup, and a
filled Fermi sphere of the complimentary atoms.  The lifetime of the state is
determined by the dimer relaxation rates $\beta_{1(23)}$ and $\beta_{3(12)}$,
which are equal by the $SU(2)$ symmetry.  We can estimate $\beta_{1(23)}$ by
extrapolating the expression for the dimer relaxation rate $\beta_{1(23)}$ in
Eq.~(\ref{beta-a23}) to 731~G, which gives $3 \times 10^{-10}$~cm$^3$/s. We
can also estimate the mean-field energy of an atom of type 1 in a (23) dimer
condensate with number density $ n_{(23)}$ by extrapolating the expression for
the atom-dimer scattering length $a_{1(23)}$ in Eq.~(\ref{aAD-a23}) to
731~G. The resulting estimate is $2 \times 10^{-9}$~Hz~cm$^3\times (2 \pi
\hbar \, n_{(23)})$.  The positive sign of the real part of $a_{1(23)}$
implies that the atoms of type 1 are repelled by the (23) dimer condensate.  A
state in which the dimer condensate and the atoms are spatially separated is
therefore energetically favored over a homogeneous state.  A final candidate
for a metastable ground state is a filled Fermi sphere of Efimov trimers.  The
binding energy $E_T^{(1)}$ of the shallower Efimov trimer is predicted to be
about 190~kHz$\times (2 \pi \hbar)$ at 731~G. Its width $\Gamma_T^{(1)}$
is predicted to be about 6~kHz$\times (2 \pi \hbar)$ at 731~G, which
corresponds to a lifetime $\hbar/\Gamma_T^{(1)}$ of about $3 \times
10^{-5}$~s.

\section{Summary}
\label{sec:discussion}

Systems consisting of $^6$Li atoms with 3 spin states provide a rich
playground for the interplay between few-body physics and many-body physics.
The experimental study of many-body physics is only possible if the loss rates
of atoms from few-body processes are sufficiently low.  Measurements of the
position and width of a single Efimov loss feature can be used to determine
the 3-body parameters $\kappa_*$ and $\eta_*$.  Calculations in the zero-range
limit can then be used to predict few-body reaction rates in the entire
universal region.

In the low-field region for $^6$Li atoms, the universal predictions were only
qualitatively successful.  
A fit to the measurements of the 3-body recombination rate 
by Ottenstein et al.\ gives the 3-body parameters 
$\kappa_* \approx 77~a_0^{-1}$ and $\eta_* \approx 0.11$.   
With these parameters, the universal results for the 3-body recombination rate 
as a function of the magnetic field
give a good fit to the narrow loss feature near 130~G but 
do not agree well with measurements in the upper half of the low-field
region \cite{Braaten:2008wd,NU:2009,Schmidt:2008fz}.  A reasonable explanation
was proposed by Wenz et al.\ \cite{Wenz:2009}: $\eta_*$ 
is particularly sensitive
to the binding energies of the shallowest of the deep dimers and there are
$^6$Li dimers whose binding energies change dramatically across the low-field
region.  Assuming the scaling behavior $\eta_* \sim E_{\rm deep}^{-1}$, they
obtained a good fit to the recombination rate in the low-field region.

In the high-field region for $^6$Li atoms, the scattering lengths are much
larger so the universal predictions should be much more accurate.
A narrow 3-atom loss feature near 895~G was discovered 
by Williams et al.\ \cite{Williams:2009} and by Jochim and coworkers
\cite{Jochim}.
By fitting their measurements of the 3-body recombination rate,
Williams et al.\ determined the 3-body parameters associated with
Efimov physics to be $\kappa_* = 80.7 \pm 2.3~a_0^{-1}$ and 
$\eta_* = 0.016^{+0.006}_{-0.010}$.  
We used those parameters to calculate
the universal predictions for the binding energies 
and widths of the Efimov
trimers shown in Fig.~\ref{fig:energieshi} and \ref{fig:widthshi}. 
The Efimov trimer responsible
for the narrow loss feature is predicted to disappear 
through the 1+(23) atom-dimer threshold at $672 \pm 2$~G, 
producing a spectacular atom-dimer loss resonance.
There is also a deeper Efimov trimer whose binding frequency 
and width in the universal region are approximately 
30~MHz and 2~MHz$\times(2 \pi \hbar)$, respectively.  
This trimer is also predicted to disappear 
through the 1+(23) atom-dimer threshold, but
this happens outside the universal region.  
We also used the 3-body parameters determined by Williams et al.\ 
to calculate the universal predictions for the
3-body recombination rate, which are shown in Figs.~\ref{fig:Khigh} and
\ref{fig:Khighlo}.  Local minima in $K_3$ are predicted at 
$759 \pm 1$~G, $829 \pm 1$~G, and $861 \pm 1$~G.
Finally an approximate calculation of the dimer relaxation rate
in the region of the atom-dimer resonance is presented in Fig.~\ref{fig:beta}.
We look forward to the experimental verification of these predictions.

In order to understand the behavior of atom-dimer mixtures at low
temperatures, it would be useful to have universal predictions for other
3-atom observables.  They include the atom-dimer scattering lengths, which are
in general complex.  The real part of an atom-dimer scattering length
determines the mean-field shifts of the atom in the dimer condensate.  Its
imaginary part is proportional to the dimer relaxation rate. For the shallow
dimer with the largest binding energy, which is the (23) dimer for $B < 730$~G
and the (12) dimer for $730~{\rm G} <B < 834$~G, the only relaxation channels
are into deep dimers.  For shallow dimers with smaller binding energy, there
are also relaxation channels into other shallow dimers.  It would be
especially useful to have definitive universal predictions for the relaxation
rate constant $\beta_{1(23)}$ near the predicted 
atom-dimer threshold at $672 \pm 2$~G.  
It would improve upon the approximation
illustrated in Fig.~\ref{fig:beta} by taking into account the $B$-dependence
of the ratios of scattering lengths. 

There have not yet been any direct observations of Efimov trimers in 
ultracold atoms.  They have only been observed indirectly through the 
resonant enhancement of 3-body recombination and through the 
resonant enhancement of atom-dimer relaxation 
provided by  virtual Efimov trimers.
The direct production of Efimov trimers would be another milestone 
in the study of Efimov physics in ultracold atoms.
Of course, once they are produced, 
they would decay quickly.  In the high-field universal region for $^6$Li atoms,
the deeper Efimov trimer is predicted to have a very short lifetime
of about $10^{-7}$~s.  
The shallower Efimov trimer is predicted to have 
a lifetime of $10^{-4}$~s to  $10^{-5}$~s.  
Our universal predictions for the 
binding energies of these Efimov trimers should be useful in devising 
experimental strategies for producing them.

\begin{acknowledgments}
We thank S.~Jochim and K.M.~O'Hara for useful communications. 
This research was supported in part by the
DOE under grants DE-FG02-05ER15715 and DE-FC02-07ER41457,
by a joint grant from AFOSR and ARO,
by the 
NSF under grant PHY-0653312,
by the BMBF under contracts 06BN411 and 06BN9006,
and by the Alexander von Humboldt Foundation.
\end{acknowledgments}


\begin{thebibliography}{99}


\bibitem{IKS08}
{\it Ultracold Fermi Gases}, ed.\
M.\ Inguscio, W.\ Ketterle, and C.\ Salomon 
(IOS Press, Amsterdam, 2008).

\bibitem{Bedaque:2006ii}
P.F.~Bedaque and J.P.~D'Incao,
Annals Phys.\ {\bf 324}, 1763 (2009)
[arXiv:cond-mat/0602525].

\bibitem{PMT06}
T.~Paananen, J.-P.~Martikainen, and P.~T\"orm\"a,
Phys.\ Rev.\ A {\bf 73}, 053606 (2006)
[arXiv:cond-mat/0603498].

\bibitem{HJZ06}
L.~He, M.~Jin, and P.~Zhuang,
Phys.\ Rev.\ A {\bf 74}, 033604 (2006)
[arXiv:cond-mat/0604580].

\bibitem{Zhai06}
H.~Zhai,
Phys.\ Rev.\ A {\bf 75}, 031603(R) (2007)
[arXiv:cond-mat/0607459].

\bibitem{MKTP09}
J.-P.~Martikainen, J.~Kinnunen, P.~Torma, and C.J.~Pethick,
arXiv:0908.3733.

\bibitem{Efimov}
V.~Efimov,
Phys.\ Lett.\ {\bf 33B}, 563 (1970).

\bibitem{Braaten:2004rn}
  E.~Braaten and H.-W.~Hammer,
  Phys.\ Rept.\  {\bf 428}, 259 (2006)
[arXiv:cond-mat/0410417].


\bibitem{Braaten:2006vd}
  E.~Braaten and H.-W.~Hammer,
  Annals Phys.\  {\bf 322}, 120 (2007)
  [arXiv:cond-mat/0612123].

\bibitem{Platter:2009gz}
  L.~Platter,
Few-Body Syst. {\bf 46}, 139 (2009), 
  [arXiv:0904.2227].

\bibitem{Efimov73}
V.~Efimov,
Nucl.\ Phys.\ A {\bf 210}, 157 (1973).

\bibitem{Kraemer:2006}
T.~Kraemer, M.~Mark, P.~Waldburger, J.G.~Danzl, C.~Chin, B.~Engeser,
A.D.~Lange, K.~Pilch, A.~Jaakkola, H.-C.~N\"agerl, and R.~Grimm,
Nature {\bf 440}, 315 (2006)
[arXiv:cond-mat/0611629].

\bibitem{NM-99}
E.~Nielsen and J.H.~Macek,
Phys.\ Rev.\ Lett.\ {\bf 83}, 1566 (1999).

\bibitem{EGB-99}
B.D.~Esry, C.H.~Greene, and J.P.~Burke,
Phys.\ Rev.\ Lett.\ {\bf 83}, 1751 (1999).

\bibitem{BBH-00}
P.F.~Bedaque, E.~Braaten, and H.-W.~Hammer,
Phys.\ Rev.\ Lett.\ {\bf 85}, 908 (2000)
        [arXiv:cond-mat/0002365].


\bibitem{Braaten:2003yc}
 E.~Braaten and H.-W.~Hammer,
 Phys.\ Rev.\  A {\bf 70}, 042706 (2004)
 [arXiv:cond-mat/0303249].

\bibitem{Knoop:2008}
S.~Knoop, F.~Ferlaino, M.~Mark, M.~Berninger, H.~Schoebel, 
H.-C.~N\"agerl, and R.~Grimm, 
Nature Physics {\bf 5}, 227 (2009) [arXiv:0807.3306].

\bibitem{Helfrich:2009uy}
 K.~Helfrich and H.-W.~Hammer,
 Europhys.\ Lett.\  {\bf 86}, 53003 (2009)
 [arXiv:0902.3410].

\bibitem{Hammer:2006ct}
  H.-W.~Hammer and L.~Platter,
  Eur.\ Phys.\ J.\  A {\bf 32}, 113 (2007)
  [arXiv:nucl-th/0610105].

\bibitem{Stecher:2008}
J.~von Stecher, J.~P.~D'Incao, and C.~H.~Greene,
Nature Physics  {\bf 5}, 417 (2009) [arXiv:0810.3876].

\bibitem{Ferlaino:2009} 
F.~Ferlaino, S.~Knoop, M.~Berninger, W.~Harm, J.~P.~D'Incao, H.-C. N{\"a}gerl,
R.~Grimm,
Phys.\ Rev.\ Lett.\ {\bf 102}, 140401 (2009) [arXiv:0903.1276].

\bibitem{Zaccanti:2008} 
M.~Zaccanti, B.~Deissler, C.~D'Errico, M.~Fattori, M.~Jona-Lasinio, 
S.~M\"uller, G.~Roati, M.~Inguscio, and G.~ Modugno,
Nature Physics {\bf 5}, 586 (2009) [arXiv:0810.3876].

\bibitem{Barontini:2009} 
G.~Barontini, C.~Weber, F.~Rabatti, J.~Catani, G.~Thalhammer, 
	M.~Inguscio, and F.~Minardi,
Phys.\ Rev.\ Lett.\ {\bf 103}, 043201 (2009) [arXiv:0901.4584].

\bibitem{Gross:2009}
N.~Gross, Z.~Shotan, S.~Kokkelmans, L.~Khaykovich,
Phys.\ Rev.\ Lett.\ {\bf 103}, 163202 (2009)
[arXiv:0906.4731].

\bibitem{Bartenstein:2005}
M.~Bartenstein, A.~Altmeyer, S.~Riedl, R.~Geursen, S.~Jochim, C.~Chin,
	J.~Hecker Denschlag, R.~Grimm,
A.~Simoni, E.~Tiesinga, C.J.~Williams, and P.S.~Julienne,
Phys.\ Rev.\ Lett.\ {\bf 94}, 103201 (2005)
[arXiv:cond-mat/0408673].

\bibitem{Ottenstein:2008}
T.B.~Ottenstein, T.~Lompe, M.~Kohnen, A.N.~Wenz, and S.~Jochim,
Phys.\ Rev.\ Lett.\ {\bf 101}, 203202 (2008) [arXiv:0806.0587].

\bibitem{Huckans:2008fq}
  J.~H.~Huckans, J.~R.~Williams, E.~L.~Hazlett, R.~W.~Stites and K.~M.~O'Hara,
  Phys.\ Rev.\ Lett.\  {\bf 102}, 165302 (2009)
  [arXiv:0810.3288].

\bibitem{Braaten:2008wd}
 E.~Braaten, H.-W.~Hammer, D.~Kang and L.~Platter,
 Phys.\ Rev.\  Lett.\ {\bf 103}, 073202 (2009)
 [arXiv:0811.3578].

\bibitem{NU:2009}
P.~Naidon and M.~Ueda,
Phys.\ Rev.\  Lett.\ {\bf 103}, 073203 (2009)
[arXiv:0811.4086].

\bibitem{Schmidt:2008fz}
 R.~Schmidt, S.~Floerchinger, and C.~Wetterich,
 Phys.\ Rev.\  A {\bf 79}, 053633 (2009)
 [arXiv:0812.1191].

\bibitem{Williams:2009} 
J.R.~Williams, E.L.~Hazlett, J.H.~Huckans, R.W.~Stites, Y.~Zhang, 
and K.M.~O'Hara, 
arXiv:0908.0789.

\bibitem{Jochim}
S.~Jochim, private communication.

\bibitem{Petrov03}
D.S.~Petrov,
Phys.\ Rev.\ A {\bf 67}, 010703(R) (2003)
[arXiv:cond-mat/0209246].

\bibitem{DIncao05}
J.P.~D'Incao and B.D.~Esry,
Phys.\ Rev.\ Lett.\ {\bf 94}, 213201 (2005)
[arXiv:cond-mat/0411565].

\bibitem{STM57}
G.V.~Skorniakov and K.A.~Ter-Martirosian,
Sov.\ Phys.\ JETP {\bf 4}, 648 (1957)
[J.\ Exptl.\ Theoret.\ Phys.\ (U.S.S.R.) {\bf 31}, 775 (1956)].

\bibitem{Hammer:2000nf}
  H.-W.~Hammer and T.~Mehen,
  Nucl.\ Phys.\  A {\bf 690}, 535 (2001)
  [arXiv:nucl-th/0011024].

\bibitem{Braaten:2001ay}
  E.~Braaten, H.-W.~Hammer and T.~Mehen,
  Phys.\ Rev.\ Lett.\  {\bf 88}, 040401 (2002) [arXiv:cond-mat/0108380].

\bibitem{Gogolin:2008}
A.O.~Gogolin, C.~Mora, and R.~Egger,
  Phys.\ Rev.\ Lett.\  {\bf 100}, 140404 (2008).

\bibitem{Macek:2005}
J.H.~Macek, S.~Ovchinnikov, and G.~Gasaneo,
Phys.\ Rev.\ A {\bf 72}, 032709 (2005).

\bibitem{Petrov-octs}
D.~Petrov,
talk at the Workshop on Strongly Interacting Quantum Gases,
Ohio State University, April 2005.

\bibitem{Petrov04}
D.S.~Petrov, C.~Salomon, and G.V.~Shlyapnikov,
Phys.\ Rev.\  Lett.\ {\bf 93}, 090404 (2004)
[arXiv:cond-mat/0309010].

\bibitem{Julienne}
P.S.~Julienne, private communication (2009).

\bibitem{Wenz:2009}
A.N.~Wenz, T.~Lompe, T.B.~Ottenstein, F.~Serwane, G.~Z\"urn, and S.~Jochim,
Phys.\ Rev.\ A {\bf 80}, 040702 (2009)
[arXiv:0906.4378].

\bibitem{Braaten-Kang}
 E.~Braaten and D.~Kang, 
 in preparation.
   


\end{thebibliography}
\end{document}